\documentclass[twocolumn]{IEEEtran}

\usepackage{cite}
\usepackage[cmex10]{amsmath}
\usepackage{graphicx}
\usepackage[caption=false,font=footnotesize]{subfig}
\usepackage{amssymb}
\usepackage{multirow}
\usepackage{algorithm}
\usepackage{algcompatible}

\DeclareMathOperator{\maximize}{maximize}
\newtheorem{prop}{Proposition}

\newtheorem{myremark}{Remark}

\allowdisplaybreaks

\begin{document}

\title{Optimal Joint User Association and Resource Allocation in Heterogeneous Networks via Sparsity Pursuit}

\author{Quan Kuang,
        Wolfgang Utschick,
        and~Andreas Dotzler

\thanks{Q. Kuang, W. Utschick and A. Dotzler are with the Department of Electrical and Computer Engineering, Technische Universit\"{a}t M\"{u}nchen, Munich 80333, Germany (e-mail: quan.kuang@tum.de; utschick@tum.de; dotzler@tum.de)}.}

\maketitle

\begin{abstract}

This paper studies the joint user association and resource allocation in heterogeneous networks (HetNets) from a novel perspective, motivated by and generalizing the idea of fractional frequency reuse. By treating the multi-cell multi-user resource allocation as resource partitioning among multiple \emph{reuse patterns}, we propose a unified framework to analyze and compare a wide range of user association and resource allocation strategies for HetNets, and provide an optimal benchmark for network performance. The enabling mechanisms are a novel formulation to consider \emph{all} possible interference patterns or \emph{any} pre-defined subset of patterns, and efficient sparsity-pursuit algorithms to find the solution. A notable feature of this formulation is that the patterns remain fixed during
the resource optimization process. This creates a favorable opportunity for
convex formulations while still considering interference coupling. More importantly, in view of the fact that multi-cell resource allocation is very computational demanding, our framework provides a systematic way to trade off performance for the reduction of computational complexity by restricting the candidate patterns to a small number of \emph{feature} patterns. Relying on the sparsity-pursuit capability of the proposed algorithms, we develop a practical guideline to identify the feature patterns. Numerical results show that the identified feature patterns can significantly improve the existing strategies, and jointly optimizing the user association and resource allocation indeed brings considerable gain.

\end{abstract}


\section{Introduction}

The heterogeneous network (HetNet), where low-power low-complexity base-stations (BSs) are overlaid with conventional macro BSs, is being considered as a promising paradigm for increasing system capacity and coverage in a cost-effective way. However, the inter-cell interference (ICI) potentially introduced by hierarchical layering of cells becomes a fundamental limiting factor to the HetNet performance. One way to control the ICI is to allocate the time/frequency resources in an intelligent manner across multiple cells, determining which BSs should transmit on which channels, and at which time instance.

 This resource allocation problem in HetNets is highly coupled with the user association policy. In a macro-only cellular network, the user association can be decoupled from the resource allocation and the user is simply associated to the BS with strongest downlink signal strength. In a HetNet, however, this association policy will lead to the case where the macro cells become resource constrained while the small cells are extremely underutilized, due to the large discrepancy in their transmit power. To balance the load, a user may be connected to a small cell even though the received power from a macro BS is higher. However, this may cause severe interference and overwhelm the cell-splitting gain eventually, if the radio resources are not carefully partitioned among cells. Clearly, the resource allocation and user association should be optimized jointly \cite{Fooladivanda2013}.

This paper studies the joint user association and multi-cell resource allocation from a novel perspective, motivated by and generalizing the idea of fractional frequency reuse (FFR) \cite{Boudreau2009,Saquib2013}. The basic mechanism of FFR is to pre-define a set of \emph{reuse patterns}. Each of these patterns determines a particular combination of ON/OFF activities of all BSs. For example, the reuse-1 pattern simply activates all the BSs, and a reuse-3 pattern activates one BS among neighboring three BSs.
 Our framework generalizes this idea by developing mechanisms to consider \emph{any} pre-defined or \emph{all possible} patterns.


 In the existing works that deal with joint user association and interference management (see for example, \cite{Shen2014, Luo1, Luo2, Luo3, Kuang2012}), power/beamforming parameters are included as design variables. Consequently, the interference coupling leads to the inherent non-convexity in the optimization problems even for the fixed user association. Actually, it has been proved in \cite{Luo2} that for a wide range of utility functions (including the proportional fairness utility in this paper), the joint user association and power allocation is NP-hard if the number of tones is larger than two.

In this paper, we develop a new framework for multicell resource allocation. The idea is to characterize the interference by pre-defining interference patterns, and then perform the resource partitioning among those patterns. An important feature of this pattern formulation is that the power parameters are specified by each pattern and remain fixed during the resource optimization. This creates an excellent opportunity for convex
formulations, while still taking into account the interference coupling.


Using the proposed framework, we develop a unified view on a wide range of existing resource management strategies, which is done by restricting the candidate patterns to a certain set of pre-defined ones. This view enables us to analyze and compare various strategies -- that have been proposed in the existing literature in very diverse contexts and seem not comparable at first glance -- in a unified manner. Since the formulated resource optimization problem is convex, it can be easily solved by interior point methods using standard solvers if the number of pre-defined patterns is small.

On the other hand, computing an optimal benchmark requires all possible patterns, resulting in exponential growth in number of variables.
Fortunately, as will show in this paper, only a small number of \textit{feature patterns} need to be considered for resource allocation without compromising the optimality.
However, unlike in the conventional homogeneous network where the mixture of reuse-1 and reuse-3 patterns is a natural choice, the selection of feature patterns in a HetNet is no longer a simple task, due to the irregular cell location and overlaid cell deployment. The proposed framework is able to identify the appropriate feature patterns for the given network.



 Specifically, we firstly formulate the network utility maximization problem by adapting the user association and multi-pattern resource allocation. Although the problem is nonconvex combinatorial, we are able to develop efficient algorithms (Algorithms 1 and 2) to compute tight upper bounds using convex relaxation. We then propose an algorithm (Algorithm 3) based on alternating optimization that is shown to achieve nearly optimal solutions to the original nonconvex combinatorial problem. 
By considering all possible $2^B-1$ interference patterns ($B$ is the number of cells), we compute an optimal benchmark
to quantify how well the existing approaches perform. Moreover, the proposed algorithms are able to find sparse solutions in the sense of activating only a small number of patterns out of $2^B-1$. Relying on this sparsity-pursuit capability, we develop practical guidelines to identify the feature patterns in the given HetNet. Finally, we compare various user association and resource allocation strategies using the proposed framework, where we show the identified feature patterns can significantly improve the existing strategies.


\subsection{Related work}

FFR has attracted lots of research efforts from both academia \cite{Chang2009, Novlan2010, Dotzler2010, Ali2009} and industrial standardizations \cite{R1-0513412005}. These frequency reuse schemes have been implemented in a static manner during the network planning phase \cite{Saquib2013, Novlan2010, Dotzler2010}, or adaptively according to the time-variations in cell traffic loads \cite{Chang2009, Ali2009}. The joint power and frequency allocation for OFDMA FFR has also been studied, for example in \cite{Lopez-Perez2012b} and references cited therein.

On the other hand, there has been an increasing interest in user association problems. While the earlier works and some recent ones focus on the minimization of the total transmit power \cite{Yates1995, Hanly1995, Rashid-Farrokhi1997, Qian2013}, the current trends are to optimize user association for load balancing in HetNets as described in \cite{Ye2013, Shen2014}. Meanwhile, simple "range expansion" techniques have been introduced in Third Generation Partnership Project (3GPP) standardization bodies to off-load macro users to small cells by adding a positive bias to the downlink signal strength of small BSs during the cell selection \cite{Madan2010}. The off-loading effect of range expansion techniques has been extensively studied, for example, in \cite{Madan2010, Guvenc2011, Guvenc2011a}.

 However, despite their coupled nature, the joint user association and frequency allocation for FFR has been less explored. In \cite{Son2009}, the authors studied a dynamic user association problem in a FFR network. However, the frequency partitioning is assumed given beforehand and fixed. By assuming that all cells use all spectrum (reuse-1 pattern), \cite{Ye2013} investigated a joint user association and \emph{intra-cell} resource allocation problem. Our formulation generalizes their formulation by taking into account the \emph{inter-cell} resource allocation as well. In particular, we study multi-pattern resource allocation as an effective mean of reducing the inter-cell interference, together with the user association strategy and intra-cell resource allocation.

 In \cite{Fooladivanda2013}, the joint multi-cell channel allocation and user association was studied. However, their study was restricted to three pre-defined resource allocation strategies, namely, orthogonal deployment, co-channel deployment, and partially shared deployment. Our framework generalizes their studies in that it allows us to consider \emph{any} pre-defined or \emph{all possible} strategies (i.e., reuse patterns) when the user association is also optimized.
 More importantly, the proposed formulation provides a vehicle to identify the essential strategies (reuse patterns) that give better performance than those defined in \cite{Fooladivanda2013} by intuition.

 The joint user association and interference coordination via time-domain almost blank subframes (ABSs) was investigated in \cite{Ye2013a, Jin2013}, where macro BSs do not transmit any data over certain subframes periodically to configure ABSs such that pico cells can schedule vulnerable users on those resources with reduced interference. Each macro BS is assumed to have the same blank subframes in \cite{Ye2013a, Jin2013}. As will become clear, this strategy can be easily analyzed using our framework by defining only two patterns: all-ON and only-pico-ON. Moreover, the proposed framework covers the more general case where each macro BS may have different blank subframes.



 In our preliminary work \cite{Kuang}, a joint user association and reuse pattern selection problem was formulated where each BS serves its associated users by a round-robin scheduler, allocating all its spectrum to a single user at a given time. Since the resulting problem is nonconvex and combinatorial, a heuristic approach based on Tabu search was adopted to solve the problem. However, it is not clear how far the solution is from the optimum. In this paper, we obtain a convex formulation for more general assumptions and the global solution is found to the formulated problem. Part of the results in this paper have been presented in a workshop \cite{wsa2015}.

 In an independent work \cite{Binnan2015}, all possible reuse patterns were also considered for the spectrum allocation in HetNets, where the average packet sojourn time is minimized taking into account the traffic variation. User association was assumed predetermined and fixed in this work.

 Note that reuse pattern selection has been studied in a \emph{Time Division Multiple Access (TDMA)} macro-only network in \cite{Son2011}, which is different from our formulation. Additionally, the HetNet deployment requires a new criterion to select essential patterns.



\subsection{Outline of the paper}

 Section II introduces the system model and problem formulation. The relaxed problem is studied in Section III, where we present the properties of the relaxed problem and the details of the proposed sparsity-pursuit algorithms for solving the problem. In Section IV, we develop an alternating algorithm to solve the original problem. In Section V, we study the feature pattern identification using our sparsity-pursuit algorithms, and compare various strategies using our framework.

\section{System model and problem formulation}

\subsection{System model}
We consider a downlink HetNet, where a number of small cells are embedded in the conventional macro cellular network\footnote{Cell and BS are used interchangeably in this paper.}.  The set of all cells is denoted as $\mathcal{B}$ with the cardinality $B = |\mathcal{B}|$. A set of users $\mathcal{K}$, with $K = |\mathcal{K}|$,  is distributed in the network, and each user is associated with only one serving cell, referred to as single-BS association.

 The enabling mechanism is to characterize the interference by specifying the interference patterns, each of which defines a particular ON/OFF combinations of BSs. We use the pattern activity vector $\boldsymbol{t}_i=(t_{i1}, t_{i2}, \cdots,t_{iB})^T$ to indicate the ON/OFF of the BSs under pattern $i$, where $t_{ib} =1$ if BS $b$ is ON; otherwise $t_{ib} =0$.

 If pattern $i$ is the only pattern in the network and user $k$ is the only user associated with BS $b$, the ergodic rate of user $k$ would be written as
\begin{equation}\label{ExclusiveRatePerCarrier}
   r_{kbi} = W \mathbb{E}_{\textbf{h}} \left[\log_2 \left(1+ \frac{t_{ib}P_b G_{kb} \|h_{kbn}\|^2}{\sigma^2 + \sum_{l\neq b} t_{il}P_l G_{kl} \|h_{kln}\|^2}\right) \right]   
\end{equation}
where $W$ is the system bandwidth, $\sqrt{G_{kb}}h_{kbn}$ is the channel gain from BS $b$ to user $k$ at channel $n$ with $G_{kb}$ accounting for path loss, shadowing and $h_{kbn}$ representing the small-scale fading, $\{h_{kbn}, \forall k, \forall b, \forall n\}$ are assumed independent and identically distributed, $\textbf{h} \triangleq (h_{k1n},h_{k2n},\ldots, h_{kBn})$, $P_b$ is the transmit power spectral density (PSD) of BS $b$, and $\sigma^2$ is the received noise PSD.

 However, the network generally has more than one interference pattern and each BS serves more than one user. The actual user rate thus depends on the resource allocation strategy. By predefining a set of candidate patterns $\mathcal{I}=\{1,2,\cdots,I\}$, $\{r_{kbi}\}$ can be pre-calculated using (\ref{ExclusiveRatePerCarrier}) and treated as constants. We can then partition the channels across these patterns.
 In our model, channels are generally referred to as orthogonal resource units in the time or/and frequency domain. 

 Specifically, let $\boldsymbol{\pi} \triangleq (\pi_1, \ldots, \pi_i, \ldots, \pi_I) \in \Pi$ be the allocation profile, where $\pi_i$ represents the fraction of total resources allocated to pattern $i$ and $\Pi = \{\boldsymbol{\pi} : \sum_i \pi_i = 1, 0 \leq \pi_i \leq 1, \forall i\}$.  We further denote by $\alpha_{kbi} \geq 0$ the fraction of resources that BS $b$ allocates to user $k$ under pattern $i$. Naturally, we have
 \begin{equation}\label{BSresourceConst}
 \sum_{k\in \mathcal{K}_b} \alpha_{kbi} \leq \pi_i, \quad \forall b,\forall i
 \end{equation}
 where $\mathcal{K}_b$ represents the set of users associated with BS $b$. To enforce single-BS association for user $k$, a \emph{binary} association indicator $a_{kb}$ is further introduced,
 i.e., $a_{kb} = 1$ if user $k$ is associated with BS $b$; otherwise it is 0. Hence, $\mathcal{K}_b = \{ k \in \mathcal{K} : a_{kb} = 1\}$ and the single-BS association requires $\sum_{b\in \mathcal{B}} a_{kb} = 1, \forall k$.
Finally we can express the average user rate after resource allocation as
\begin{equation}\label{RateRestricted}
  R_k  =  \sum_{b \in \mathcal{B}} \sum_{i \in \mathcal{I}}  a_{kb}\alpha_{kbi} r_{kbi}.
\end{equation}

\begin{myremark}
The ergodic rate is used in our system model because of two reasons. First, the adaptation of user association and multi-cell resource assignment is expected to perform over a large time scale, e.g., in minutes or hours. Too frequent changes of user association and multi-cell resource allocation will introduce severe overhead and delay issues and hence deteriorate the user experience. So it is impractical to adapt according to the fast fading channel state information (in milliseconds). Second, the ergodic rate formulation results in equal channel conditions among all frequency resources, which enables a unified treatment for either time or frequency allocation such that the user rate $R_k$ scales linearly with the assigned resources $\alpha_{kbi}$ as shown in (\ref{RateRestricted}).
\end{myremark}

\begin{myremark}
In this paper, we aim at developing strategies for long-term user association and multi-cell resource allocation based on the average channel information. On top of this adaptation, each cell can perform individual channel-aware scheduling for all its associated users among the agreed spectrum in a more frequent manner to respond to fast fading channel fluctuations. Our design serves the first layer in this two-level adaptation.
\end{myremark}

\begin{myremark}
In practical systems, such as LTE/LTE-A, the frequency subcarriers are grouped into resource blocks that are the minimum units for resource allocation. However, our formulation assumes resource allocation can be performed in a continuous way without granularity limit, which bounds the performance in reality. It is a reasonable approximation when the system has a large number of resource units.
\end{myremark}

\subsection{Problem formulation}\label{section_2.2}
Our objective is to maximize the long-term network utility by adapting the user association and multi-pattern resource allocation, which is formulated as
\begin{IEEEeqnarray}{rCl}\label{problem_rst}
    \displaystyle\mathop{\maximize}_{\boldsymbol{\alpha},\boldsymbol{\pi}, \boldsymbol{a}}
                \quad && U = \sum_{k \in \mathcal{K}} \ \omega_k \log(R_k)   \IEEEyessubnumber \label{rst_obj} \\
    \text{subject to} \quad && R_k =  \sum_{b \in \mathcal{B}} \sum_{i \in \mathcal{I}} a_{kb} \alpha_{kbi} r_{kbi} \IEEEyessubnumber  \\
                      && \sum_{k \in \mathcal{K}} a_{kb} \alpha_{kbi} \leq \pi_i, \ \forall b,i   \IEEEyessubnumber \label{rst_const_BS allo}\\
                      && \sum_{b \in \mathcal{B}} a_{kb} = 1, \ \forall k \IEEEyessubnumber \label{cons_asso} \\
                      && a_{kb} \in \{0,1\} \IEEEyessubnumber \label{cons_integer} \\
                      && \sum_{i \in \mathcal{I}} \pi_i = 1 \IEEEyessubnumber \label{rst_cons_pi} \\
                      && \pi_i \geq 0, \forall i, \quad \alpha_{kbi} \geq 0, \forall k,b,i   \IEEEyessubnumber \label{rst_con_nonnegative}
\end{IEEEeqnarray}
where a logarithmic utility function is used because it establishes a very good balance between network throughput and fairness among the users and hence is commonly used in existing studies (e.g., \cite{Ye2013, Shen2014, Fooladivanda2013}), the weights $\omega_k$ provide a means for service differentiation, and the BS resource constraints in (\ref{rst_const_BS allo}) are equivalent to (\ref{BSresourceConst}). Note that our framework also applies to any other concave differentiable utility functions.

The problem (\ref{problem_rst}) is a combinatorial nonconvex problem, hence difficult to solve. Before presenting our algorithm to solve it, we investigate a convex relaxation in the next section, which provides a performance upper bound. Interestingly, as we will show by numerical results, the bound is actually very tight. 

\section{Multi-BS association: a relaxed problem}


\subsection{Relaxed problem formulation}\label{section_3.1}
By dropping the binary association indicator, we arrive at:
\begin{IEEEeqnarray}{rCl}\label{problem_convex}
    \displaystyle\mathop{\maximize}_{\boldsymbol{\alpha},\boldsymbol{\pi}}
                \quad && U = \sum_{k \in \mathcal{K}} \ \omega_k \log(R_k)   \IEEEyessubnumber \label{obj} \\
    \text{subject to} \quad && R_k = \sum_{i \in \mathcal{I}} \sum_{b \in \mathcal{B}}  \alpha_{kbi} r_{kbi} \IEEEyessubnumber  \\
                      && \sum_{k \in \mathcal{K}} \alpha_{kbi} \leq \pi_i, \forall b,\ \forall i   \IEEEyessubnumber \label{const_BS allo}\\
                      && \sum_{i \in \mathcal{I}} \pi_i = 1 \IEEEyessubnumber \label{cons_pi} \\
                      && \pi_i \geq 0, \forall i \quad \alpha_{kbi} \geq 0, \forall k,b,i   \IEEEyessubnumber \label{con_nonnegative}
\end{IEEEeqnarray}
which is a relaxation of problem (\ref{problem_rst}). To see this, let $a_{kb}, \alpha_{kbi}, \pi_i$ be any feasible point of (\ref{problem_rst}). We define $\alpha^\prime_{kbi} = a_{kb}\alpha_{kbi}$ and $\pi^\prime_i = \pi_i$. It can be easily verified that $\alpha^\prime_{kbi}$ and $\pi^\prime_i$ are feasible with respect to problem (\ref{problem_convex}). In other words, the feasible region of problem (\ref{problem_rst}) is a subset of that of problem  (\ref{problem_convex}). Since the two problems optimize the same objective, (\ref{problem_convex}) is a relaxation of (\ref{problem_rst}).

In (\ref{problem_convex}), the user association is implicitly indicated by $\alpha_{kbi}$, i.e., $\alpha_{kbi} > 0$ means user $k$ is associated with BS $b$ under pattern $i$, while zero value of $\alpha_{kbi}$ means that they are not connected. In this sense, user $k$ is allowed to be connected to multiple BSs, as long as the summation of the resources allocated to all users does not exceed the allowable value (see (\ref{const_BS allo})) at each BS over each pattern. In the following, we term this relaxation \emph{multi-BS association}.

 The formulation of (\ref{problem_convex}) is a convex optimization problem since we maximize a concave function over a convex set. It can be efficiently solved by, e.g., interior point methods, using standard solvers, if the number of candidate patterns is small. On the other hand, we would also like to solve (\ref{problem_convex}) by considering all possible $2^B-1$ patterns in the network, which will provide
 an optimal benchmark for comparison with other existing approaches where pattern selection is restricted. In this case, the computational complexity is polynomial in $2^B$ using off-the-shelf interior-point solvers. It becomes infeasible for a reasonable-sized network. In the following, we will present two efficient algorithms tailored to the problem at hand, resulting in linear worst-case complexity in $2^B$. Furthermore, the proposed algorithms have the ability to select a few essential patterns out of $2^B-1$ to achieve the optimality.

\begin{myremark}
The formulation of (4) in \cite{Ye2013} can be regarded as a special case of the proposed formulation (\ref{problem_convex}) by letting reuse-1 be the only allowable pattern. The proposed formulation introduces multi-pattern resource allocation as another degree of freedom for system design.
\end{myremark}

\begin{myremark}
The investigation in \cite{Ye2013a} can also be cast into the proposed framework. Specifically, by restricting the candidate patterns to two with partially shared resources, our formulation boils down to the optimization problem formulated in (5) in \cite{Ye2013a}. Being more general, the proposed formulation can be used to investigate the case where each macro BS has different blank subframes.
\end{myremark}


\subsection{Some properties}





The following Propositions \ref{prop_NumPattern} and \ref{prop_NumMultiUser} reveal properties of the optimal solution to problem (\ref{problem_convex}), motivating the development of our algorithms.

 \begin{prop}\label{prop_NumPattern}
  There exists an optimal solution to convex problem (\ref{problem_convex}) that activates at most $K$ patterns in the network, i.e., $|\mathcal{I}^{\text{on}}| \leq  K$, where $\mathcal{I}^{\text{on}} = \{i \in \mathcal{I} : \pi_{ i } > 0 \}$, and $K$ is the number of users in the network.
 \end{prop}

\begin{IEEEproof}
By letting $\alpha_{kbi} = \pi_i \rho_{kbi}$, the problem of (\ref{problem_convex}) can be equivalently rewritten as
\begin{IEEEeqnarray}{rCl}\label{problem_rewrite}
    \displaystyle\mathop{\maximize}_{\boldsymbol{\rho},\boldsymbol{\pi}}
                \quad &&  U = \sum_{k \in \mathcal{K}} \ \omega_k \log(R_k)   \IEEEyessubnumber \label{obj_rewrite} \\
    \text{subject to} \quad && R_k = \sum_{i \in \mathcal{I}} \sum_{b \in \mathcal{B}}  \pi_i \rho_{kbi} r_{kbi} \IEEEyessubnumber \label{rate_constr_rewrite} \\
                      && \sum_{k \in \mathcal{K}} \rho_{kbi} \leq 1, \forall b,\ \forall i\in \mathcal{I}^{\text{on}}   \IEEEyessubnumber \label{const_BS allo_rewrite}\\
                      && \sum_{i \in \mathcal{I}} \pi_i = 1 \IEEEyessubnumber \label{cons_pi_rewrite} \\
                      && \pi_i \geq 0, \forall i, \quad \rho_{kbi} \geq 0, \forall k,b,i.   \IEEEyessubnumber \label{con_nonnegative_rewrite}
\end{IEEEeqnarray}

Let us denote the user rate vector as $\boldsymbol{R} = [R_1, \cdots, R_K]^T $. The \emph{achievable rate region} defined by the constraints from (\ref{rate_constr_rewrite}) to (\ref{con_nonnegative_rewrite}) can be expressed as the convex hull of the pattern driven regions, i.e.,
\begin{IEEEeqnarray}{rCl}\label{rateRegion}
  \mathcal{R} &&= \text{conv}(\mathcal{R}^1, \cdots, \mathcal{R}^I)
  \end{IEEEeqnarray}
where 
\begin{IEEEeqnarray}{lCl}\label{patternRateRegion}
  \mathcal{R}^i = \{ && \boldsymbol{R}^i = [R_1^i, \cdots, R_K^i]^T \ :\ R_k^i = \sum_{b \in \mathcal{B}} \rho_{kbi} r_{kbi},\IEEEnonumber\\
      &&   \sum_{k \in \mathcal{K}} \rho_{kbi} \leq 1, \forall b, \ \rho_{kbi} \geq 0, \forall k,b   \}.
\end{IEEEeqnarray}

According to the Caratheodory's theorem \cite[Theorem 2.1.6]{Bazaraa2013}, any point $\boldsymbol{R} \in \mathcal{R} \subset \mathbb{R}_{+}^K$ can be expressed as the convex combination of at most $K+1$ points in $\bigcup_{i=1}^I \mathcal{R}^i$, i.e., $\boldsymbol{R}$ lies in a $d$-simplex with vertices in $\bigcup_{i=1}^I \mathcal{R}^i$ and $d\leq K$. Hence, no more than $K+1$ patterns are needed to represent any feasible rate satisfying (\ref{rate_constr_rewrite}) to (\ref{con_nonnegative_rewrite}). Furthermore, since the optimal $\boldsymbol{R}^\star$ to (\ref{problem_rewrite}) must be Pareto efficient with respect to all the $K$ users (otherwise we can increase the objective function (\ref{obj_rewrite}) by moving to the Pareto optimum), it cannot be an interior point of the $d$-simplex, and must lie on some $m$-face of the $d$-simplex with $m<d$. Therefore, $\boldsymbol{R}^\star$ can be written  as a convex combination of at most $K$ points.
\end{IEEEproof}

An immediate consequence of Proposition \ref{prop_NumPattern} is that although the number of all possible patterns grows exponentially with the number of cells in the network, we can allocate nonzero fraction of resources to only a small number of patterns to achieve the optimality. Note that resource allocation among the set of transmission modes has been studied in a TDMA context in \cite{Son2011, Raman2005}. They conjectured that almost always only very few active transmission modes are needed as corroborated by their simulation results. Relying on 
describing the original rate region by the convex hull of the pattern driven regions, we are able to bound the cardinality of the set of active patterns theoretically, which as a byproduct validates their conjecture.

Although the objective function of (\ref{problem_convex}) is strongly concave with respect to $R_k$, it not strictly concave in $\alpha_{kbi}$ and $\pi_i$. Thus, the optimal solution is generally not unique. How to pursuit the sparse solution in the sense of activating less patterns is not answered by the Proposition \ref{prop_NumPattern}, which is the topic in the next subsection.

 \begin{prop}\label{prop_NumMultiUser}
 In optimal solutions to convex problem (\ref{problem_convex}), the number of users who are associated with multiple BSs over the same pattern or the same set of patterns is at most $B-1$, where $B$ is the number of cells in the network.
 \end{prop}
 \begin{IEEEproof}
The proof is based on the KKT conditions and provided in Appendix.
\end{IEEEproof}

The implication of Proposition \ref{prop_NumMultiUser} is remarkable. It indicates as a result of optimizing resource allocation and user association, most of users in the network will be associated with only one BS in any given pattern, although we allow multi-association in our relaxed problem formulation, hinting that the relaxation would probably be really tight.


\subsection{Proposed approach}\label{section_alg1}

Our approaches stem from the Frank-Wolfe method, also known as the conditional gradient method \cite{Bertsekas1999}. As one of the simplest first-order methods that has been known since 1950s, Frank-Wolfe-type methods have recently re-gained interest in several areas, including machine learning particularly, mainly due to its good scalability, and the crucial property of enabling sparse solutions \cite{Jaggi2013}. We present our first approach in the following Algorithm 1.

\begin{algorithm}
\label{alg1}
\caption{Multi-BS User Association and Pattern Selection}
\begin{algorithmic}[1]
\STATE Initialize iteration counter $t = 1$, tolerance $\epsilon > 0$ ; Choose $(\boldsymbol{\alpha}^1, \boldsymbol{\pi}^1) \in \mathcal{X}$;
\REPEAT
\STATE Compute $(\bar{\boldsymbol{\alpha}}, \bar{\boldsymbol{\pi}}) = \arg \max_{(\boldsymbol{\alpha}, \boldsymbol{\pi}) \in \mathcal{X}} \langle \boldsymbol{\alpha}, \nabla_{\boldsymbol{\alpha}} U(\boldsymbol{\alpha}^t) \rangle$;
\STATE Update $\boldsymbol{\alpha}^{t+1} = \boldsymbol{\alpha}^t + \gamma^t (\bar{\boldsymbol{\alpha}}- \boldsymbol{\alpha}^t)$, \     $ \boldsymbol{\pi}^{t+1} = \boldsymbol{\pi}^t + \gamma^t ( \bar{\boldsymbol{\pi}} -  \boldsymbol{\pi}^t ) $, where step size $\gamma^t \in [0,1]$ is chosen by a warm-start Armijo rule;
\STATE $t = t +1$;
\UNTIL optimality certificate is less than $\epsilon$.
\end{algorithmic}
\end{algorithm}

In Algorithm 1, we denote the feasible set defined by (\ref{const_BS allo}), (\ref{cons_pi}) and (\ref{con_nonnegative}) as $\mathcal{X}$. The gradient $\nabla_{\boldsymbol{\alpha}} U(\boldsymbol{\alpha}^t)$ is defined with the entries $[\nabla_{\boldsymbol{\alpha}} U(\boldsymbol{\alpha}^t)]_{kbi} = \frac{\partial U(\boldsymbol{\alpha}) }{\partial \alpha_{kbi}} \mid_{\boldsymbol{\alpha}=\boldsymbol{\alpha}^t}$. The inner product $\langle \boldsymbol{\alpha}, \nabla_{\boldsymbol{\alpha}} U(\boldsymbol{\alpha}^t) \rangle = \sum_{k=1}^K \sum_{b=1}^B \sum_{i=1}^I \alpha_{kbi}[\nabla_{\boldsymbol{\alpha}} U(\boldsymbol{\alpha}^t)]_{kbi}$. We explain the key elements of Algorithm 1 as follows.

\subsubsection{Solving the linear subproblem}
Solving the linear subproblem in step 3 is the most critical step for Algorithm 1. A close look reveals that it has the following explicit analytic solution:

  \begin{prop}\label{prop_solvingLinearProblem}
The solution to the linear subproblem in Algorithm 1 is

\begin{equation}\label{solu_alp}
  \bar{\alpha}_{kbi} =    \left\{  \begin{array}{rl}
                                        1 & \text{if} \ i=\bar{i}, k=\bar{k}(b, \bar{i})  \\
                                        0 & \text{otherwise}
                              \end{array} \right.
\end{equation}
and
\begin{equation}\label{solu_pi}
  \bar{\pi}_i =    \left\{  \begin{array}{rl}
                                        1 & \text{if} \ i=\bar{i}  \\
                                        0 & \text{otherwise}
                              \end{array} \right.
\end{equation}
where
\begin{equation}\label{optimal_k}
  \bar{k}(b,i) = \arg \max_k [\nabla_{\boldsymbol{\alpha}} U(\boldsymbol{\alpha}^t)]_{kbi}
\end{equation}

\begin{equation}\label{optimal_i}
  \bar{i} = \arg \max_i \sum_b[\nabla_{\boldsymbol{\alpha}} U(\boldsymbol{\alpha}^t)]_{\bar{k}(b,i)bi}.
\end{equation}

 \end{prop}

  \begin{IEEEproof}
The linear subproblem can be rewritten as the following inner-outer formulation:
\begin{IEEEeqnarray}{rCl}\label{inner_outer linear}
      \displaystyle\mathop{\maximize}_{\substack{\pi_i \geq 0 \\ \sum_i \pi_i = 1}} \quad \displaystyle\mathop{\maximize}_{\substack{\alpha_{kbi} \geq 0 \\ \sum_k \alpha_{kbi}\leq \pi_i }} \sum_{k=1}^K \sum_{b=1}^B \sum_{i=1}^I \alpha_{kbi}[\nabla_{\boldsymbol{\alpha}} U(\boldsymbol{\alpha}^t)]_{kbi}. \IEEEeqnarraynumspace
\end{IEEEeqnarray}
It is clear that the inner problem is solved by each BS exclusively allocating maximum allowable resources to one user who benefits the most for each pattern, i.e.,
\begin{equation}\label{solu_alp_inner}
  \alpha_{kbi} =    \left\{  \begin{array}{rl}
                                        \pi_i & \text{if} \ k=\bar{k}(b, i), \forall b, \forall i  \\
                                        0 & \text{otherwise}
                              \end{array} \right.
\end{equation}
where $\bar{k}$ is expressed as (\ref{optimal_k}). Substituting the solution of (\ref{solu_alp_inner}) back to (\ref{inner_outer linear}), we arrive at the following problem:
\begin{IEEEeqnarray}{rCl}\label{outer linear}
      \displaystyle\mathop{\maximize}_{\pi_i \geq 0, \sum_i \pi_i = 1}  \sum_{i=1}^I \pi_i  \sum_{b=1}^B  [\nabla_{\boldsymbol{\alpha}} U(\boldsymbol{\alpha}^t)]_{\bar{k}(b,i)bi}
\end{IEEEeqnarray}
which is solved by pooling all resources to one pattern. So we obtain the solutions of (\ref{optimal_i}) and (\ref{solu_pi}), hence (\ref{solu_alp}).
\end{IEEEproof}

\begin{myremark}
Proposition \ref{prop_solvingLinearProblem} reveals that at most one new pattern is activated in each iteration. Hence, by initializing it with single pattern, Algorithm 1 has the potential to identify sparse solutions of using at most $t$ patterns, where $t$ is the iteration counter. A notable feature of the proposed algorithm, as we will show later, is that whenever we stop the iteration, we know how far the current solution is from the global optimum by evaluating a gap function (see (\ref{duality gap})). 
\end{myremark}

\subsubsection{Choosing the step size}

The basic Armijo rule \cite{Bertsekas1999} is to set $\gamma^t = \beta^{m_t}$, where $\beta \in (0,1)$ and $m_t$ is the first nonnegative integer for which
\begin{equation}\label{Armijio}
  U(\boldsymbol{\alpha}^{t+1}) \geq U(\boldsymbol{\alpha}^{t}) + \kappa  \langle  \boldsymbol{\alpha}^{t+1}- \boldsymbol{\alpha}^t, \nabla_{\boldsymbol{\alpha}} U(\boldsymbol{\alpha}^t) \rangle
\end{equation}
with $\kappa \in (0,1)$ fixed. The idea is that we start with the initial step size 1 and continue to reduce to $\beta$, $\beta^2$, ..., until the we find the largest $\beta^{m_t}$ such that the $\kappa$ improvement of utility function by its linear estimation is achieved.

We adopt a warm-start variant of the basic Armijo rule. In view of the fact that $\gamma^{t-1}$ and $\gamma^t$ may be similar, instead of starting from 1 every time we use $\gamma^{t-1}$ as the initial guess and then either increase or decrease it in order to find the largest $\beta^{m_t}$ satisfying (\ref{Armijio}). Specifically, we initially set $\gamma^t = \gamma^{t-1}$; If the condition (\ref{Armijio}) is satisfied, we set $\gamma^t = \min(\gamma^t / \beta, 1)$ until (\ref{Armijio}) does not hold or $\gamma^t = 1$; Else repeatedly decrease $\gamma^t$ by setting it to $\beta \gamma^t$.


\subsubsection{Initialization}

Any feasible point can be used to initialize the algorithm, leading to an optimal solution. In order to obtain a sparse solution in particular, we start the algorithm with a single active pattern in the network and single-BS association for all users. For simplicity, we can always choose the reuse-1 pattern if it is in the candidate pattern set, otherwise we randomly pick one pattern from the candidates. We then allocate all resources to this selected pattern $\hat{i}$, as
\begin{equation}\label{ini_opt_pi}
    \pi^1_i =    \left\{  \begin{array}{rl}
                                        1 & \text{if} \ i=\hat{i}  \\
                                        0 & \text{otherwise.}
                              \end{array} \right.
\end{equation}
To initialize $\boldsymbol\alpha$, each user chooses the BS with largest $r_{kb\hat{i}}$ under pattern $\hat{i}$. Within each BS, resources are then uniformly distributed among its associated users as
\begin{equation}\label{ini_opt_alp}
  \alpha^1_{kbi} =    \left\{  \begin{array}{rl}
                                        \frac{1}{K_{b}} & \text{if} \ i=\hat{i} \ \text{and} \  K_{b} \neq 0   \\
                                        0 & \text{otherwise}
                              \end{array} \right.
\end{equation}
where $K_{b}$ is the number of users associated with BS $b$.

\subsubsection{Optimality certificate and convergence}
We denote by $\boldsymbol{\alpha}^\star$, $\boldsymbol{\pi}^\star$ the solution to problem (\ref{problem_convex}), i.e., $U(\boldsymbol{\alpha}^\star) \geq U(\boldsymbol{\alpha}), \forall (\boldsymbol{\alpha}, \boldsymbol{\pi}) \in \mathcal{X}$. The $\epsilon$-optimal solution set is defined as
\begin{equation}\label{optimalSetDef}
  \mathcal{S}^\star_\epsilon = \{ (\boldsymbol{\alpha}, \boldsymbol{\pi}) \in \mathcal{X} : U(\boldsymbol{\alpha}^\star)- U(\boldsymbol{\alpha}) \leq \epsilon     \}.
\end{equation}

 Due to the concavity of the objective function, we have
\begin{IEEEeqnarray}{rCl}\label{optimaltyBound}
  U(\boldsymbol{\alpha}^\star)- U(\boldsymbol{\alpha}^t) && \leq  \langle \nabla_{\boldsymbol{\alpha}} U(\boldsymbol{\alpha}^t), \boldsymbol{\alpha}^\star - \boldsymbol{\alpha}^t   \rangle \IEEEnonumber\\
   && \leq \displaystyle\mathop{\maximize}_{(\boldsymbol{\alpha}, \boldsymbol{\pi}) \in \mathcal{X}} \langle \nabla_{\boldsymbol{\alpha}} U(\boldsymbol{\alpha}^t), \boldsymbol{\alpha} - \boldsymbol{\alpha}^t   \rangle
\end{IEEEeqnarray}
where the second inequality follows because $\boldsymbol{\alpha}^\star$ is feasible.

Let us define the optimality gap function as
\begin{equation}\label{duality gap}
  g(\boldsymbol{\alpha}^t) = \displaystyle\mathop{\maximize}_{(\boldsymbol{\alpha}, \boldsymbol{\pi}) \in \mathcal{X}} \langle \nabla_{\boldsymbol{\alpha}} U(\boldsymbol{\alpha}^t), \boldsymbol{\alpha} - \boldsymbol{\alpha}^t  \rangle =  \langle \nabla_{\boldsymbol{\alpha}} U(\boldsymbol{\alpha}^t), \bar{\boldsymbol{\alpha}} - \boldsymbol{\alpha}^t   \rangle.
\end{equation}
Hence, the value of $g(\boldsymbol{\alpha}^t)$ can be easily obtained as a by-product of every iteration of Algorithm 1. If the current iteration satisfies $g(\boldsymbol{\alpha}^t) \leq \epsilon$, it is guaranteed for $\boldsymbol{\alpha}^t, \boldsymbol{\pi}^t$ being $\epsilon$-optimal (i.e., $(\boldsymbol{\alpha}^t, \boldsymbol{\pi}^t) \in \mathcal{S}^\star_\epsilon$).

 In order to use $g(\boldsymbol{\alpha}^t) \leq \epsilon$ as the stopping criterion, we need to make sure that $g(\boldsymbol{\alpha}) \rightarrow 0$ if $(\boldsymbol{\alpha}, \boldsymbol{\pi}) \rightarrow (\boldsymbol{\alpha}^\star, \boldsymbol{\pi}^\star)$. According to optimality condition of a differentiable concave function, the optimal solution should satisfy  \cite{Boyd2004}:
\begin{equation}\label{optimalCondition}
  \langle \nabla_{\boldsymbol{\alpha}} U(\boldsymbol{\alpha}^\star), \boldsymbol{\alpha} - \boldsymbol{\alpha}^\star   \rangle \leq 0, \  \forall (\boldsymbol{\alpha}, \boldsymbol{\pi}) \in \mathcal{X}
\end{equation}
which means
\begin{equation}\label{cvgCondition}
  \displaystyle\mathop{\maximize}_{(\boldsymbol{\alpha}, \boldsymbol{\pi}) \in \mathcal{X}} \langle \nabla_{\boldsymbol{\alpha}} U(\boldsymbol{\alpha}^\star), \boldsymbol{\alpha} - \boldsymbol{\alpha}^\star   \rangle = g(\boldsymbol{\alpha}^\star)= 0.
\end{equation}
As a result, we have the following convergence result:

\begin{prop}\label{convergence}
$(\boldsymbol{\alpha}^t, \boldsymbol{\pi}^t)$ generated by Algorithm 1 converges to $\epsilon$-optimal solution.
\end{prop}

 \begin{IEEEproof}
 Proposition 2.2.1 in \cite{Bertsekas1999} suggests that every limit point of $\{\boldsymbol{\alpha}^t\}$ is $\boldsymbol{\alpha}^\star$. Together with (\ref{cvgCondition}), we have $g(\boldsymbol{\alpha}^t) \rightarrow 0 $ as $t \rightarrow \infty$.
 If we stop the algorithm after a finite number of steps and condition $g(\boldsymbol{\alpha}^t) \leq \epsilon$ is met, the solution is in the $\epsilon$-optimal solution set according to the definition (\ref{optimalSetDef}).
 \end{IEEEproof}


\subsection{Variant algorithm for enhanced sparsity}
\label{section_alg2}

\begin{algorithm}
\caption{Sparsity-Enhanced Pattern Selection and Multi-BS User Association}
\begin{enumerate}
  \item[] ... as Algorithm 1, except replacing step 4 with
  \item[\footnotesize{4:}] Let $\mathcal{I}^{t} = \{ \bar{i} \}$ where $\bar{i}$ is obtained from step 3 (see (\ref{optimal_i})).   \\     Update $(\boldsymbol{\alpha}^{t+1}, \boldsymbol{\pi}^{t+1})$ by solving problem (\ref{problem_convex}) using interior point methods with $\mathcal{I} =\mathcal{I}^{1}\bigcup \mathcal{I}^{2} \cdots \bigcup \mathcal{I}^{t}$.
\end{enumerate}
\end{algorithm}

Algorithm 2 is a "fully corrective" variant of the Algorithm 1. In each iteration, after a new pattern is identified, we re-optimize the original problem over all previously identified patterns. Compared to Algorithm 1, more progress is made per iteration. Therefore, less number of iterations and hence better sparsity are expected by Algorithm 2. Since we solve the original problem over a significantly reduced dimension (considering $t$ instead of $2^B-1$ patterns in the $t$-th iteration), the complexity of interior point methods is not an issue any more. Thanks to the sparsity nature of optimal solutions as identified by Proposition \ref{prop_NumPattern}, the Algorithm 2 normally converges within $K$ iterations. In Section V, we will compare the two algorithms with respect to the sparsity pursuit capability.

 \begin{myremark}
If we solve (\ref{problem_convex}) using either Algorithm 1 or 2 when considering all $2^B-1$ patterns, the complexity will be dominated in step 3 that is linear in $2^B$ (since we are looking for the largest value among $2^B-1$ values), as $K$ and $B$ increase.
\end{myremark}

\section{Single-BS association}

In the previous section, we have solved a relaxed problem where user can have multiple associations and the associations can be different under different patterns. We now focus on the original problem formulated in (\ref{problem_rst}) that only allows single-BS association.

\subsection{Outline of the algorithm}

 Our approach is to leverage algorithms proposed in the previous section. Specifically, we group the variables into two blocks, $(\boldsymbol{\alpha},\boldsymbol{\pi})$ and $\boldsymbol{a}$, and optimize them alternatively with the other block fixed as shown in Algorithm 3. The details and interesting characteristics of Algorithm 3 are provided in the following Section \ref{section 4.2}.

\begin{algorithm}
\label{alg2}
\caption{Single-BS User Association and Pattern Selection}
\begin{algorithmic}[1]
\STATE \textbf{Initialization}: $a_{kb} = 1, \forall k, \forall b$ ;
\REPEAT
\STATE Solve (\ref{problem_rst}) for fixed $\boldsymbol{a}$ by Algorithm 1 or Algorithm 2;
\STATE Solve (\ref{problem_rst}) for fixed $(\boldsymbol{\alpha}, \boldsymbol{\pi})$;
\UNTIL objective function cannot be increased.
\end{algorithmic}
\end{algorithm}

\subsection{Algorithm details and characteristics}\label{section 4.2}

\subsubsection{Pattern allocation for fixed user association}\label{section_fixeduser asso}

We first show that the problem (\ref{problem_rst}) with fixed association $\boldsymbol{a}$ can be solved exactly by Algorithm 1 or Algorithm 2. By writing $\mathring{r}_{kbi} = a_{kb} r_{kbi}$, we rewrite (\ref{problem_rst}) to the following problem
\begin{IEEEeqnarray}{rCl}\label{subproblem_fixed_a}
    \displaystyle\mathop{\maximize}_{\boldsymbol{\alpha},\boldsymbol{\pi}}
                \quad && U = \sum_{k \in \mathcal{K}} \ \omega_k \log(R_k)   \IEEEyessubnumber \label{sub_rst_obj} \\
    \text{subject to} \quad && R_k = \sum_{i \in \mathcal{I}} \sum_{b \in \mathcal{B}}  \alpha_{kbi} \mathring{r}_{kbi} \IEEEyessubnumber  \\
                      && \sum_{k \in \mathcal{K}_b}  \alpha_{kbi} \leq \pi_i, \forall b,\ \forall i   \IEEEyessubnumber \label{sub_rst_const_BS allo}\\
                      && \sum_{i \in \mathcal{I}} \pi_i = 1 \IEEEyessubnumber \label{sub_rst_cons_pi} \\
                      && \pi_i \geq 0, \quad \alpha_{kbi} \geq 0   \IEEEyessubnumber \label{sub_rst_con_nonnegative}
\end{IEEEeqnarray}
which is almost the same as the relaxed problem (\ref{problem_convex}) with the only two differences. First, the summation in the constraint (\ref{sub_rst_const_BS allo}) is over $\mathcal{K}_b$. Second, the problem (\ref{subproblem_fixed_a}) is solved when the effective rate $\mathring{r}_{kbi} \triangleq  a_{kb} r_{kbi}$, instead of $r_{kbi}$, are given. As (\ref{problem_convex}), problem (\ref{subproblem_fixed_a}) is also convex and has the same structure. Hence, Algorithm 1 or 2 can be directly applied to solve the problem to the desired accuracy. 
Interestingly, the maximization operation in (\ref{optimal_k}) can be taken over $\mathcal{K}$ without damaging the optimality (see the following Proposition \ref{universalFeasibility}).

The following proposition states an equivalent reformulation of (\ref{subproblem_fixed_a}), which we refer to as universal feasibility property.

 \begin{prop}\label{universalFeasibility}
The problem (\ref{subproblem_fixed_a}) can be equivalently solved by replacing $\sum_{k \in \mathcal{K}_b}  \alpha_{kbi} \leq \pi_i$ with $\sum_{k \in \mathcal{K}}  \alpha_{kbi} \leq \pi_i$.
 \end{prop}

 \begin{IEEEproof}
 Let $(\boldsymbol{\alpha}^\text{new}$, $\boldsymbol{\pi}^\text{new})$ be the solution to the reformulated problem where $k \in \mathcal{K}_b$ in (\ref{sub_rst_const_BS allo}) has been replaced with $k \in \mathcal{K}$. 
 Note that $(\boldsymbol{\alpha}^\text{new}, \boldsymbol{\pi}^\text{new})$ is also feasible in the original problem (\ref{subproblem_fixed_a}), since $\mathcal{K}_b \subseteq \mathcal{K}$ and $\alpha_{kbi} \geq 0$. Next we prove that $(\boldsymbol{\alpha}^\text{new}, \boldsymbol{\pi}^\text{new})$ must be the solution to the problem (\ref{subproblem_fixed_a})  by contradiction. Suppose it is not true, meaning that we can find a feasible point in problem (\ref{subproblem_fixed_a}), say $(\boldsymbol\alpha^\text{old}, \boldsymbol\pi^\text{old})$, such that $U(\boldsymbol\alpha^\text{old}) > U(\boldsymbol\alpha^\text{new})$. Then we can construct another feasible point $(\boldsymbol\alpha^\prime, \boldsymbol{\pi}^\prime)$ by choosing $\alpha_{kbi}^\prime = \left\{  \begin{array}{rl}
                                        0 & \text{if} \ k \notin \mathcal{K}_b  \\
                                        \alpha_{kbi}^\text{old} & \text{otherwise}
                              \end{array} \right.$  and $\boldsymbol\pi^\prime = \boldsymbol\pi^\text{old}$, respectively,
resulting in the same objective, i.e., $U(\boldsymbol\alpha^\prime) = U(\boldsymbol\alpha^\text{old})$, since $\mathring{r}_{kbi} = 0$ if $k\notin \mathcal{K}_b, \forall b, \forall i$. Note that $(\boldsymbol\alpha^\prime,  \boldsymbol{\pi}^\prime)$ is also feasible in the reformulated problem and $U(\boldsymbol\alpha^\prime) > U(\boldsymbol\alpha^\text{new})$, which is contradictory to the optimality of $(\boldsymbol{\alpha}^\text{new}, \boldsymbol{\pi}^\text{new})$.
\end{IEEEproof}

 \begin{myremark}
 The universal feasibility identified by Proposition \ref{universalFeasibility} will significantly simplify the calculation of user association update as given in the following subsection, by removing the coupling constraints.
 \end{myremark}

\subsubsection{User association update}

After we define $R_{kb} = \sum_{i} \alpha_{kbi} r_{kbi}$, the problem (\ref{problem_rst}) with fixed $(\boldsymbol{\alpha}, \boldsymbol{\pi})$ can be rewritten as
\begin{IEEEeqnarray}{rCl}\label{subproblem_fixedAlphaPi}
    \displaystyle\mathop{\maximize}_{\boldsymbol{a}}
                \quad && U = \sum_{k \in \mathcal{K}} \ \omega_k \log(R_k)   \IEEEyessubnumber  \\
    \text{subject to} \quad && R_k = \sum_{b \in \mathcal{B}}  a_{kb} R_{kb} \IEEEyessubnumber  \\
                      && \sum_{b \in \mathcal{B}} a_{kb} = 1, \forall k \IEEEyessubnumber  \\
                      && a_{kb} \in \{0,1\} \IEEEyessubnumber .
\end{IEEEeqnarray}
Note that constraints in (\ref{rst_const_BS allo}) have been dropped because they are automatically satisfied if we solve (\ref{subproblem_fixed_a}) using the universal feasibility constraints as given by Proposition \ref{universalFeasibility}.

It can be easily seen that problem (\ref{subproblem_fixedAlphaPi}) can be decoupled into $K$ problems, which can be solved exactly by each user choosing the best BS that give the largest
$R_{kb}$, i.e.,
\begin{equation*}
  a^\circ_{kb} =  \left\{  \begin{array}{rl}
                                        1 & \text{if} \ b \in b^\circ(k)  \\
                                        0 & \text{otherwise}
                              \end{array} \right.
\end{equation*}
where $\ b^\circ(k) = \arg\max_{b\in \mathcal{B}} R_{kb} $.

\subsubsection{Initialization and convergence}

A natural choice of initialization is the unrestricted user association, i.e., $a_{kb} = 1, \forall k, \forall b$. Although it is not a feasible $\boldsymbol{a}$ in terms of the single-BS association constraints, it produces the upper bound after applying Algorithm 1 or 2, and then becomes feasible after one iteration of Algorithm 3. It beats random initialization significantly in our experiments.

Since both the pattern allocation update and user association update in Algorithm 3 maximize the same utility function, the overall algorithm produces nondecreasing objective function values. The whole algorithm is guaranteed to converge because the utility function is finite. Although the converged solution is not necessarily global optimal, we can always bound the performance loss by comparing it to the upper bound that is obtained by solving the relaxed problem in Section III. As we will show in the numerical results, the Algorithm 3 can achieve nearly optimal solution.

\section{Performance evaluation and discussions}

\subsection{Simulation scenarios and parameter setting}

\begin{figure}[!t]
\centering
\includegraphics[width=3.5In]{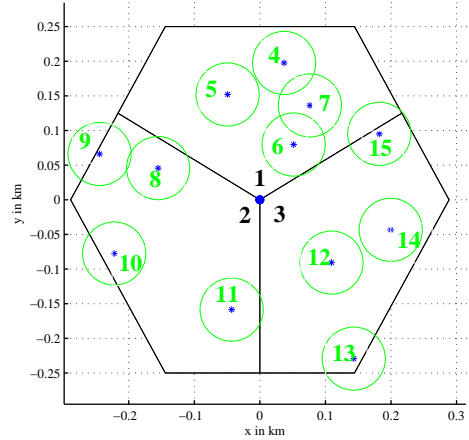}
\caption{A heterogeneous network consisting of 15 cells. }\label{fig_network}
\end{figure}

\begin{table}[!t]
\renewcommand{\arraystretch}{1.0}
\caption{Network parameters.}
\label{table1}
\centering
\begin{tabular}{c c}
\hline\hline 
Parameter & Description \\ [0.5ex] 
\hline 
bandwidth & 10 MHz  \\ 
Macro total Tx power   & 46 dBm  \\
Pico total Tx power    & 30 dBm \\
Macro antenna gain & 15 dB\\
Pico antenna gain & 5 dB \\
Macro path loss & $128.1+37.6\log_{10}(R)$ \\
Pico path loss & $140.7+36.7 \log_{10}(R)$ \\
Penetration loss & 20 dB \\
Shadowing std. dev. & 8dB(macro), 10dB(pico) \\
Shadowing corr. distance & 25 m \\
Macrocell shadowing corr. & 1 between cells \\
Picocell shadowing corr. & 0.5 between cells \\
Fading model & No fast fading \\
Min. macro(pico)-UE dist. & 35 m (10 m) \\
Min. macro(pico)-pico dist. & 75 m (40 m)\\
Noise density and noise figure & -174 dBm/Hz, 9dB\\
 \hline\hline 
\end{tabular}
\end{table}

We consider a network consisting of 3 macro cells, each of which contains 4 randomly dropped pico cells as shown in Fig.\ref{fig_network}. The cells are labelled as
\begin{equation*}
  \underbrace{1,2,3}_{\text{macro cells}}, \underbrace{4,5,6,7}_\text{picos in cell 1}, \underbrace{8,9,10,11}_\text{picos in cell 2}, \underbrace{12,13,14,15}_\text{picos in cell 3}.
\end{equation*}

The parameters for propagation modelling and simulations follow the suggestions in 3GPP evaluation methodology \cite{3GPP2010}, and summarized in Table \ref{table1}. In the area under consideration, we choose user equipment (UE) density of around 420 and 225 active UEs/sq-km (corresponding to dense urban and urban environment \cite{Deb2014}), resulting in $K=90$ and 50 respectively. All UEs have unit weights ($\omega_k = 1$). In simulation, the total number of users are uniformly distributed in the network, and we average over 5 drops of users and pico locations for each UE density.

There are only two parameters regarding the step size rule, $\gamma^0$ and $\beta$, which need to be set before Algorithm 1 is run, and the performance is insensitive to their values. In the simulation, we set initial step size $\gamma^0 = 10^{-4}$ and $\beta = 0.8$. In Algorithm 2, the reduced-size subproblem is solved by cvx, a package for specifying and solving convex programs \cite{cvx}. For the optimality tolerance (see (\ref{optimalSetDef})), we set $\epsilon = 1$ for low-UE-density case and $\epsilon = 2$ for the high-UE-density case, where user rates are in bit/s.

\subsection{Tightness of the multi-BS relaxation}

Table~\ref{tableTight} compares the utility (i.e., sum of log of user rate) achieved by Algorithms 1, 2 and 3, where user rates are in bit/s. Since Algorithm 3 can use either Algorithm 1 or 2 as its subroutine, we include both cases in Table~\ref{tableTight}. To exploit the maximum freedom in pattern allocation, we consider all $2^{15}-1$ patterns in the test network.
 \begin{table}[t]
\renewcommand{\arraystretch}{1.0}
\caption{Utility values for proposed algorithms after convergence. }
\label{tableTight}
\centering
\begin{tabular}{|c | c | c | c |c| }
\hline
 &  \begin{tabular}[x]{@{}c@{}}Alg 1 \\ (multi-BS)\end{tabular} &  \begin{tabular}[x]{@{}c@{}}Alg 2 \\ (multi-BS)\end{tabular}   &  \begin{tabular}[x]{@{}c@{}}Alg 3 + Alg 1 \\ (single-BS)\end{tabular} & \begin{tabular}[x]{@{}c@{}}Alg 3 + Alg 2 \\ (single-BS)\end{tabular} \\
\hline
50 UEs &  769.90   & 770.09   & 769.79 & 770.00  \\
\hline
90 UEs &  1339.32  &  1339.60  & 1339.19 & 1339.52  \\
\hline 
\end{tabular}
\end{table}

The first observation is that Algorithms 1 and 2 converges to $\epsilon$-optimal solutions with slightly different objective values. Nevertheless, according to our value of $\epsilon$, all the obtained solutions are guaranteed to achieve more than $99.85\%$ global optimum in both UE-density cases. We also observe that single-BS association achieves almost the same as the multi-BS association. This has two important indications. It verifies that our multi-BS relaxation provides a very tight upper bound. It also shows that our Algorithm 3 for solving the nonconvex mixed-integer problem is almost optimal because it achieves nearly the upper bounds promised by the convex relaxation.

\subsection{Feature pattern identification }\label{section_featurePatt}

As the number of all possible patterns in the network grows exponentially with the number of cells, it is necessary to identify the most important patterns and allocate resources only to these feature patterns in order to reduce the complexity of the algorithm. As described in Sections \ref{section_alg1} and \ref{section_alg2}, the proposed Algorithms 1 and 2 have sparsity-pursuit capability to activate only a small number of patterns in the final solution. In this subsection, we focus on the feature pattern identification with the aid of Algorithms 1 and 2.



In \cite{Fooladivanda2013}, several strategies of selecting reuse patterns in HetNet have been proposed by intuition. However, the question whether there exists better strategies remains open. We answer this question by considering \emph{all} possible patterns in the test network, and rely on the sparsity-pursuit capability of the proposed algorithms to identify the feature patterns. Our approach provides a systematic way to find the \emph{best} reuse strategies. Table~\ref{table2} and Table~\ref{table3} list the number of active patterns after Algorithms 1 and 2 converges to $\epsilon$-optimal solution ($\geq 99.85\%$ global optimum), respectively. Recall that pattern $i$ is referred to as active if $\pi_i > 0$ in the final solution. As shown, both algorithms are effective in finding sparse solutions, activating only a small number of patterns out of $2^{15}-1$. Moreover, Algorithm 2 produces better sparsity due to more progress per iteration as explained in Section \ref{section_alg2}.

 \begin{table}[t]
\renewcommand{\arraystretch}{1.0}
\caption{Number of active patterns after Algorithm 1 converges. }
\label{table2}
\centering
\begin{tabular}{|c | c | c | c |c | c| }
\hline
 &  Drop 1  &  Drop 2   &  Drop 3  &  Drop 4  &  Drop 5 \\
\hline
50 UEs &   53 & 62 & 46 & 53 & 37 \\
\hline
90 UEs &  94 & 127  & 78 & 73 & 107 \\
\hline 
\end{tabular}
\end{table}

 \begin{table}[t]
\renewcommand{\arraystretch}{1.0}
\caption{Number of active patterns after Algorithm 2 converges. }
\label{table3}
\centering
\begin{tabular}{|c | c | c | c |c | c| }
\hline
 &  Drop 1  &  Drop 2   &  Drop 3  &  Drop 4  &  Drop 5 \\
\hline
50 UEs & 15   &  22 & 12  & 14 & 16 \\
\hline
90 UEs & 20  &  22 & 22 & 19 & 19 \\
\hline 
\end{tabular}
\end{table}

Obviously, the set of active patterns generated by Algorithm 1 (Algorithm 2) are different in each drop, varying in accordance to the user distribution, pico distribution, fading environment, etc. After examining all the results we tested, we propose the following general guideline for feature pattern selection in a HetNet \emph{without} resorting to the Algorithm 1 or 2.

\begin{itemize}
  \item All macros are OFF, and all picos are ON.
  \item One macro is ON among the adjacent three macros, and all picos are ON except those in the active macro cells.
\end{itemize}

The principle can be summarized as macro-OFF-pico-ON policy. One advantage of this policy is that the resulting set of candidate patterns is very small. For example, in our test network, there are only 4 candidate patterns needed by applying this guideline. A detailed performance comparison is provided in the next subsection.


%

\subsection{Comparing various strategies}

In this subsection, we illustrate how to use our framework as a unified way to compare various existing user association and resource allocation schemes.

\subsubsection{Comparative schemes}

The six schemes that we compare are as follows, all of which can be computed by the proposed Algorithm 3.
\begin{itemize}
  \item \emph{All-Pattern}: All $2^{15}-1$ patterns are included in the candidate pattern set.
  \item \emph{Preselected Feature Patterns (Fea-Pattern)}: The candidate patterns are restricted to four preselected ones according to our guideline proposed in Section \ref{section_featurePatt}. Specifically, if we denote a pattern by the set of \emph{muted} BSs under the pattern, these four patterns are $\{1,2,3\}$, $\{2,3,4,5,6,7\}$, $\{1,3,8,9,10,11\}$, and $\{1,2,12,13,14,15\}$.
  \item \emph{Orthogonal Deployment with Reuse-1 (OD-1)} : This is one of the strategies studied in \cite{Fooladivanda2013}, where the macro layer and pico layer are deployed on the different frequency channels, so there is no inter-layer interference. The intra-layer interference is handled by a reuse factor. In this \emph{OD-1}, reuse-1 is simply used to allow maximum bandwidth at each BS. In \cite{Fooladivanda2013}, the channel allocation between macro and pico layers was solved, together with user association, by exhaustive search, in view of the fact that the set of channels in the system is discrete and finite. Here we study this strategy using our framework. In particular, this \emph{OD-1} deployment can be regarded as restricting the candidate patterns in our framework to two, which mute all macro and all pico cells respectively. Namely, $\{1,2,3\}$ and $\{4,5,6,7,8,9,10,11,12,13,14,15\}$ are the only two allowed patterns.
  \item \emph{Orthogonal Deployment with Pico Reuse-3 (OD-3)} : This is another strategy investigated in \cite{Fooladivanda2013}, which is similar to \emph{OD-1}. The only difference is that the pico cells share the frequency channels allocated to the pico layer by a reuse factor of three. To study \emph{OD-3} using our framework, we simply restrict the candidate patterns to the following four by muting: $\{4,5,6,7,8,9,10,11,12,13,14,15\}$, $\{1,2,3,5,6,8,9,11,12,14,15\}$, $\{1,2,3,4,6,7,9,10,12,13,15\}$, and $\{1,2,3,4,5,7,8,10,11,13,14\}$, respectively.
  \item \emph{Synchronous Blank Subframes in Macro Tier (Macro ABS)} : This is the time domain enhanced intercell interference coordination (eICIC) scheme studied in \cite{Ye2013a}. We can easily investigate this strategy using our Algorithm 3, by defining two candidate patterns, muting $\{\emptyset\}$ and \{1,2,3\}, respectively.
  \item \emph{Reuse-1}: By assuming reuse-1, user association problem has been studied in \cite{Ye2013}, which is a special case of our formulation by restricting candidate pattern to one single pattern that activating all cells (i.e., muting $\{\emptyset\}$).
\end{itemize}

 We summarize how we study various schemes using our framework in Table~\ref{tableSchemes}.

 \begin{table*}[t]
\renewcommand{\arraystretch}{1.0}
\caption{Schemes for comparison. }
\label{tableSchemes}
\centering
\begin{tabular}{|c |c | c | c | c |c | c| }
\hline
  &  All-Pattern  &  Fea-Pattern   &  OD-1  &  OD-3  &  Macro ABS & Reuse-1 \\
\hline
Candidate patterns &  all $2^{15}-1$  &  4 by guideline & 2 by definition & 4 by definition& 2 by definition & 1 by definition \\
\hline
User association rule & \multicolumn{6}{|c|}{Single-BS} \\
\hline
Solving algorithm & \multicolumn{6}{|c|}{Algorithm 3}  \\

\hline 
\end{tabular}
\end{table*}

\subsubsection{Performance metrics}

The following performance metrics are considered in our comparison.

\begin{itemize}
  \item Geometric mean of user throughput, defined as $\sqrt[K]{\prod_{k=1}^K R_k}$. Maximizing the geometric mean throughput is equivalent to maximizing our utility with $\omega_k = 1$.
  \item Total sum rate of the system, defined as $\sum_{k=1}^K R_k$.
  \item Cumulative Distribution Function (CDF) of the user throughput.
\end{itemize}

The reason why we are interested in more metrics than objective function is that they are key performance indicators (KPIs) for network operators. We want to show how different user association and resource allocation schemes impact on these KPIs.

\subsubsection{Results}

\begin{figure}%
\centering
\parbox{3in}{\includegraphics[width=3.4In]{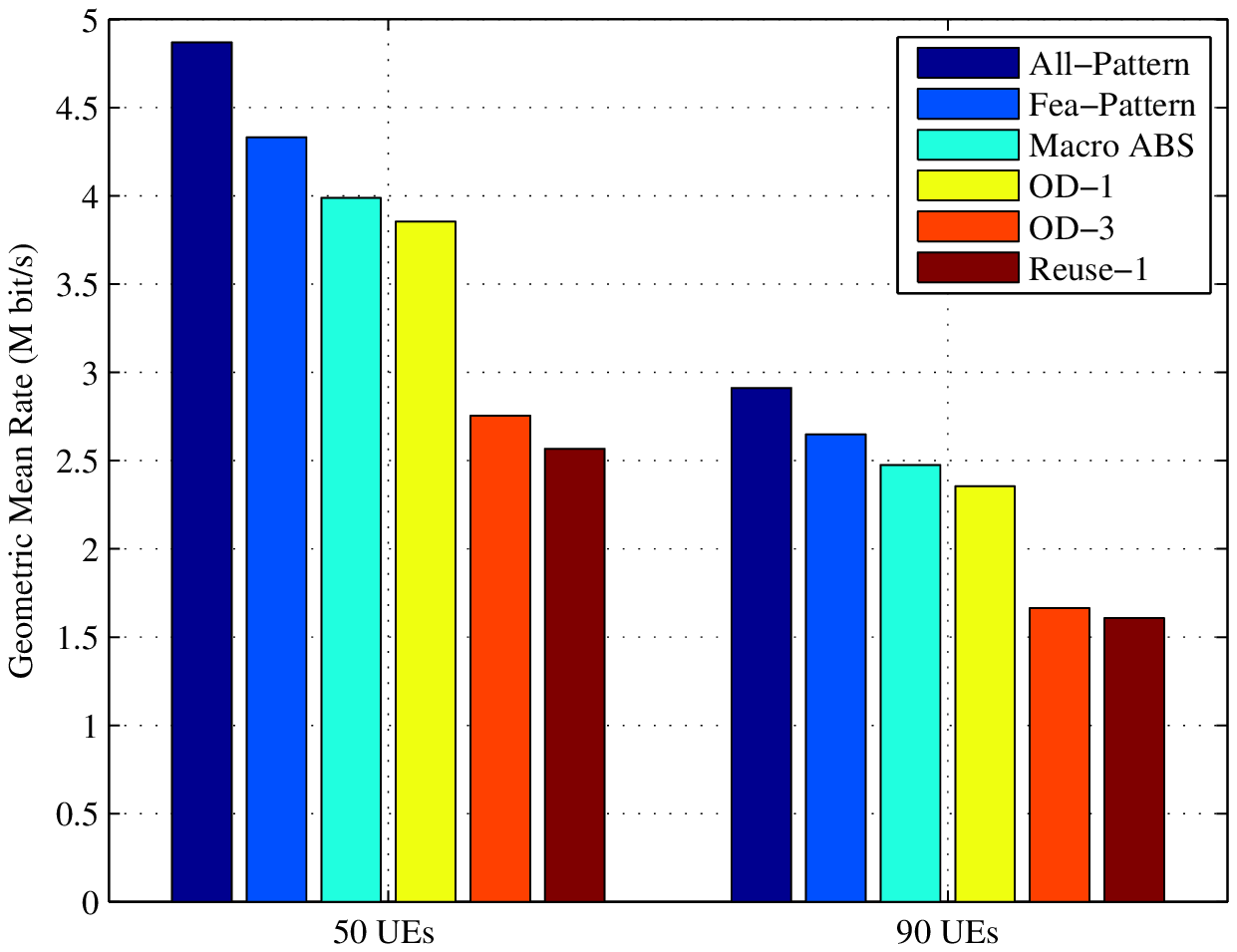}}%
\quad
\begin{minipage}{3in}%
\includegraphics[width=3.4In]{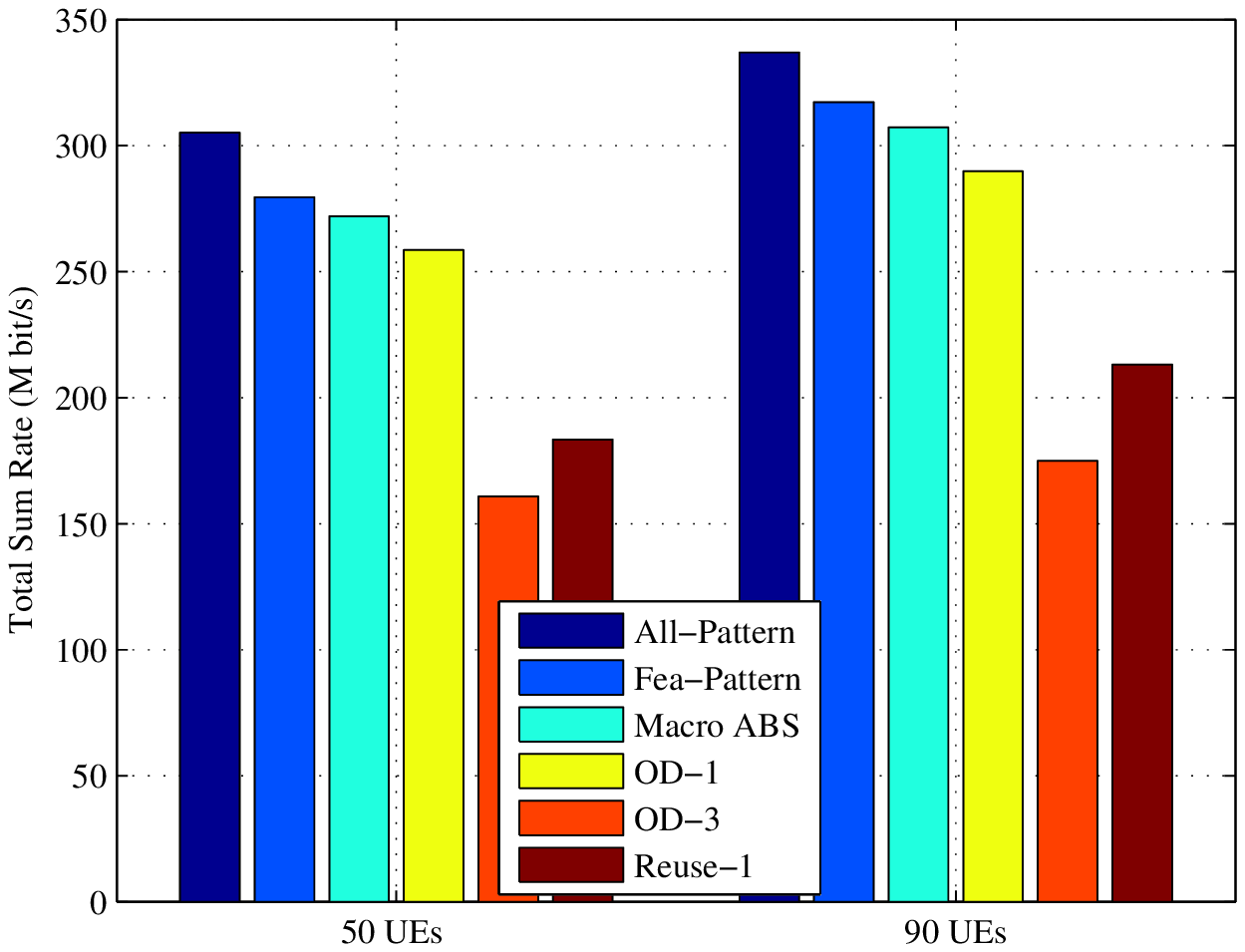}
\end{minipage}%
\caption{Geometric mean of user rate, and total system sum rate with different schemes.}%
\label{fig_bar1}%
\end{figure}

Fig.~\ref{fig_bar1} reports the geometric mean of user rate and total system sum rate for all schemes in comparison. As shown, \emph{All-Pattern} scheme indeed provides the best performance, since it has the maximum degree of freedom for pattern allocation. \emph{Fea-Pattern} scheme that relies on the proposed practical guideline to identify the feature patterns is very effective, achieving $89\%$ and $91\%$ of the \emph{All-Pattern} in terms of the geometric mean rate in 50-UE and 90-UE cases respectively, and $92\%$ and $94\%$ in terms of the sum rate. This is remarkable because the optimization is only performed over the four selected candidate patterns, requiring much less computational complexity in comparison to \emph{All-Pattern}. It performs significantly better than the existing strategies, namely, \emph{OD-1}, \emph{OD-3}, \emph{Macro ABS}, and \emph{Reuse-1}.

Among those existing strategies, \emph{Macro ABS} performs the closest to our \emph{Fea-Pattern}, immediately followed by \emph{OD-1}. This is because these two existing methods happen to share the similar spirit of our pattern selection guideline by treating the umbrella macro as the dominant interferer to all the pico cells within its coverage and defining one pattern to mute this strongest interferer. The advantage of \emph{Macro ABS} over \emph{OD-1} indicates that partially sharing resources between macro and pico tiers is more efficient than orthogonal deployment.


\emph{OD-3} suggests to deploy reuse-3 among all picos, which turns out to be inefficient as evident in Fig.~\ref{fig_bar1}. The reason is that only limited interference among pico BSs due to the low transmit power. Including pico reuse-3 patterns reduces the available resources at each pico BS and hence damages the system sum rate as shown in the second panel of Fig.~\ref{fig_bar1}.

As expected, \emph{Reuse-1} obtains the worst geometric mean rate since there is no mechanism for inter-cell interference management. In terms of the sum rate, however, it outperforms \emph{OD-3}. This is because the cell-center users who are less affected by inter-cell interference can benefit from the access to full resources provided by \emph{Reuse-1}. We make this point clear by providing the CDF of user throughputs in the next.

\begin{figure}[!t]
\centering
\includegraphics[width=3.5In]{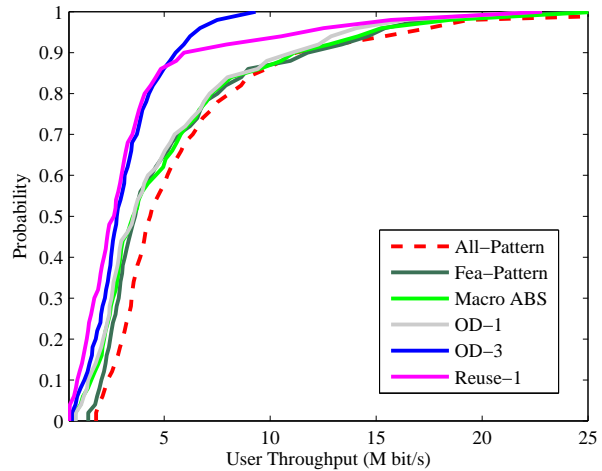}
\caption{CDF of user throughputs with different schemes for 50 UEs in the network}\label{fig_cdf50UE}
\end{figure}

\begin{figure}[!t]
\centering
\includegraphics[width=3.5In]{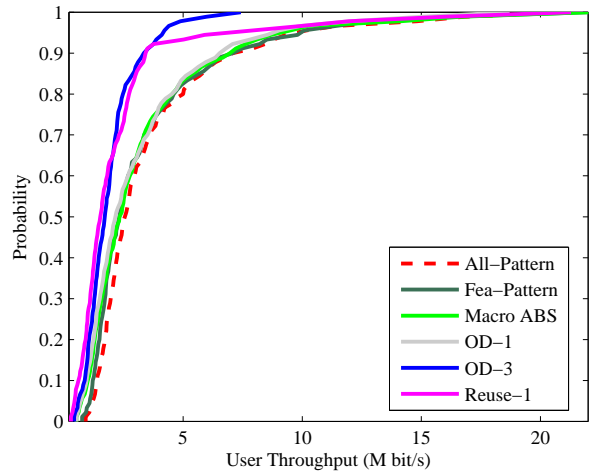}
\caption{CDF of user throughputs with different schemes for 90 UEs in the network}\label{fig_cdf90UE}
\end{figure}

Figs.~\ref{fig_cdf50UE}  and \ref{fig_cdf90UE} show the empirical CDF of user throughputs for 50 UEs and 90 UEs in the network, respectively. Being consistent with results shown in Fig.~\ref{fig_bar1}, the proposed \emph{Fea-Pattern} produces the closest curve to \emph{All-Pattern}. Both \emph{Macro ABS} and \emph{OD-1} are less effective in reducing the interference compared to \emph{Fea-Pattern}, resulting in performance loss for the cell-edge users (5th percentile throughput). Moreover, \emph{OD-1} is not as good as \emph{Macro ABS} in utilizing the resources, causing further degradation for cell-center users (95th percentile throughput) compared to \emph{Macro ABS}.
As for \emph{OD-3}, the large gap between \emph{OD-3} and all the other schemes with respect to 95th percentile throughput reveals that pico reuse-3 deployment in \emph{OD-3} is a waste of system resources, causing notable degradation in sum rate as shown in the  Fig.~\ref{fig_bar1}. 
When comparing Fig.~\ref{fig_cdf50UE} and Fig.~\ref{fig_cdf90UE}, we notice that the performance gaps among different schemes reduce when number of users in the network increases. This is because less resources are available for individual users as the number of users increases while the total resources are fixed in the network. So the gains achieved by optimizing the resource allocation decrease.

\subsection{Comparison with RE association rule}

Range expansion (RE) techniques have recently been discussed in 3GPP as a simple association rule to balance the load in HetNets. In this subsection, we study the impact of this simple association on the performance if resource allocation can be optimized, using our framework. Specifically, when the user association is fixed, the resource allocation can be optimized by solving problem (\ref{subproblem_fixed_a}) as described in Section \ref{section_fixeduser asso}. In the evaluation, we set the macro bias to zero, and the same bias for all the pico BSs, choosing from 0, 5, 10, 15, 20 and 25 dB. The UE is associated with the cell with largest value of downlink received power plus bias.

\begin{figure}[!t]
\centering
\includegraphics[width=3.5In]{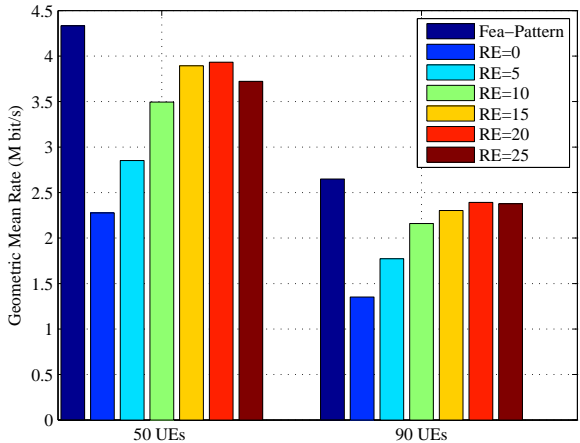}
\caption{Comparison of Fea-Pattern scheme with simple association rules based on pico range expansion, in terms of geometric mean of user rate.}\label{fig_FixedA_4pattern}
\end{figure}

Fig.~\ref{fig_FixedA_4pattern} provides the comparison of \emph{Fea-Pattern} scheme with other schemes of fixed RE association, where the resource allocation is performed over the same four feature patterns. As seen, the conventional association scheme without bias can only obtain roughly 50\% of the geometric mean rate that is achieved by the joint optimization of the \emph{Fea-Pattern} scheme in both 50-UE and 90-UE cases. However, by tuning the bias value to optimum (20 dB in the tested case), the simple RE association method with optimized resource allocation achieves 90\% of the geometric mean rate offered by \emph{Fea-Pattern}. The small gap exists because all pico BSs have to select the same bias value.

\begin{figure}[!t]
\centering
\includegraphics[width=3.5In]{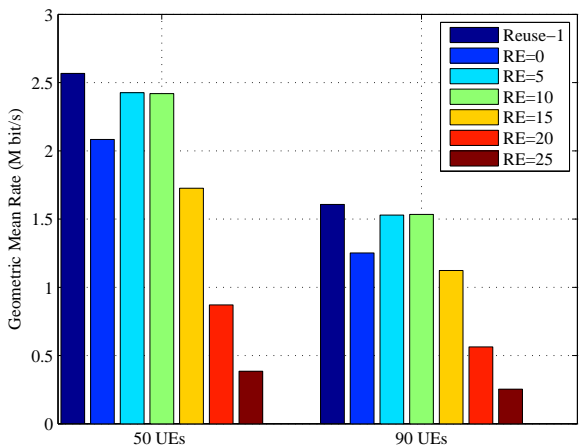}
\caption{Comparison of Reuse-1 scheme with simple association rules based on pico range expansion, in terms of geometric mean of user rate.}\label{fig_FixedA_1pattern}
\end{figure}

Fig.~\ref{fig_FixedA_1pattern} shows the similar comparison, but the resource allocation is restricted to single reuse-1 pattern. In this case, the resource allocation boils down to intra-cell resource distribution. As shown, the performance loss caused by the conventional association without bias is significantly less than that in Fig.~\ref{fig_FixedA_4pattern}. This is because reuse-1 pattern suffers from lack of inter-cell interference management, limiting the potential of joint optimization. Like in Fig.~\ref{fig_FixedA_4pattern}, the RE association with optimal bias (5 dB in Fig.~\ref{fig_FixedA_1pattern}) can achieve 95\% performance of the joint optimization. Another interesting observation is that the optimal bias value in Fig.~\ref{fig_FixedA_1pattern} is smaller than that in Fig.~\ref{fig_FixedA_4pattern}, indicating that with mechanisms to combat the inter-cell interference pico cells can use more aggressive bias value to incorporate more users. The fact that optimal bias values are different in Figs.~\ref{fig_FixedA_4pattern} and \ref{fig_FixedA_1pattern} also reveals the coupling effect of the resource allocation and user association.

The above observations coincide perfectly with the recent studies in \cite{Ye2013,AndrewsJ2014}, where a simple per-tier biasing is shown to nearly achieve the optimal performance if the bias value is chosen carefully. It is also reported in \cite{AndrewsJ2014} that the optimal bias value is considerably more aggressive in out-of-band range expansion or co-channel deployment with eICIC, in comparison with co-channel deployment without any interference coordination (see Table 1 in \cite{AndrewsJ2014}), hinting at a strong coupling between biasing and resource partitioning.

Although Figs.~\ref{fig_FixedA_4pattern} and \ref{fig_FixedA_1pattern} demonstrate the effectiveness of the RE association rule, it is, in general, difficult to prescribe the optimal biases leveraging optimization techniques \cite{AndrewsJ2014}. There are some studies based on stochastic geometry where the per tier bias parameters can be found analytically by averaging out all the potential network configurations \cite{SinghS2013}. However, it is a nontrivial task to perform this kind of optimization for the current network configuration. By contrast, in the proposed schemes we have a systematic way to optimize the user association with resource allocation.


\section{Conclusion}

In this paper, we have studied the joint user association and interference management in heterogeneous networks. We treat the multi-cell resource allocation as resource partitioning among multiple reuse patterns, and optimize the multi-cell multi-user channel assignment together with user association. The formulated problem is nonconvex and combinatorial due to the single-BS association restriction. We have provided a tight convex relaxation where multi-BS association is allowed, and proposed efficient sparsity-pursuit algorithms to find the optimal solution where only a small number of patterns are activated. The result provides an optimal upper bound of the original problem. We have also derived an efficient iterative algorithm to obtain solutions to the original problem that are close to the optimal upper bound.


An important observation is that although the number of all possible patterns grows exponentially with the number of cells in the network, most of the patterns are not used thanks to the sparsity pursuit capability of the proposed algorithms. This motivates us to restrict the candidate patterns to a set of pre-defined feature patterns in order to reduce the computation efforts. By analyzing the active patterns resulting from the proposed algorithms, we have developed practical guideline to select the feature patterns in a HetNet.

 The proposed framework enables us to compare a wide range of user association and resource allocation strategies in a unified view. Our results indicate that existing criteria for reuse pattern selection result in large performance loss in comparison to the optimal benchmark. However, the feature patterns identified by the proposed guideline significantly improves the existing strategies. We have also compared our joint optimization to the strategies where the user association is performed separately according to simple range expansion rules. It is observed that the RE rule suffers only minor performance loss, provided the bias value is chosen optimally and the resource allocation is optimized. The fact that the optimal bias values should be set differently in alignment with different resource allocation strategies reveals the coupling effect of both elements.
Only downlink is considered in this paper. Due to the downlink and uplink imbalance, the optimal downlink user association need not be optimal for the uplink transmission. A joint study of the downlink and uplink user association is an interesting topic for future research.

\appendix[Proof of Proposition 2]

Without loss of generality, we set $\omega_k = 1, \forall k$.
We define the Lagrangian associated with problem (\ref{problem_convex}) as
\begin{IEEEeqnarray}{rCl}
\label{lagragian}
L(\boldsymbol\alpha, \boldsymbol\pi, \boldsymbol\mu, \sigma) = \sum_k  \log(R_k) + \sum_i \sum_b \mu_{bi} (\pi_i - \sum_k \alpha_{kbi}) \IEEEnonumber\\
+ \sigma (1 - \sum_i \pi_i) \IEEEeqnarraynumspace
\end{IEEEeqnarray}
where $\boldsymbol\mu$ and $\sigma$ are Lagrange multipliers associated with (\ref{const_BS allo}) and (\ref{cons_pi}), respectively, the nonnegative constraints in (\ref{con_nonnegative}) are considered implicitly.

The optimal solution should satisfy the following KKT conditions:

\begin{IEEEeqnarray}{rCl}\label{KKT}
\frac{r_{kbi}}{R_k} -\mu_{bi} = 0, \quad \text{if} \ \alpha_{kbi} > 0 \IEEEyessubnumber \label{KKT_alp} \\
\sum_b \mu_{bi} = \sigma, \quad \text{if} \ \pi_i > 0 \IEEEyessubnumber \label{KKT pi} \\
\sum_k \alpha_{kbi} \leq \pi_i, \ \mu_{bi}(\sum_k \alpha_{kbi}-\pi_i) = 0 \IEEEyessubnumber \\
\mu_{bi}\geq 0, \ \alpha_{kbi} \geq 0, \ \pi_i \geq 0, \  \sum_i \pi_i = 1. \IEEEyessubnumber
\end{IEEEeqnarray}

Assume that user $\bar{k}$ is associated with two BSs, $b$ and $b'$, over the same set of patterns $\mathcal{P}$, i.e., $\alpha_{\bar{k}li} >0, l\in\{b,b'\},  i\in \mathcal{P}$.
 According to (\ref{KKT_alp}) we have

\begin{IEEEeqnarray}{rCl}
  \frac{r_{\bar{k}bi}}{R_{\bar{k} }} = \mu_{bi}, \quad \forall i\in \mathcal{P} \label{equalities1}\\
  \frac{r_{\bar{k}b'j}}{R_{\bar{k} }} = \mu_{b'j}, \quad \forall j\in\mathcal{P}. \label{equalities2}
\end{IEEEeqnarray}

For the ease of notation, we define $r_{\bar{k}b} = \sum_{i \in \mathcal{P} } r_{\bar{k}bi}$ and $\mu_b = \sum_{i \in \mathcal{P}} \mu_{bi}$. From (\ref{equalities1}) and (\ref{equalities2}), we obtain
\begin{equation}\label{cascadedEqu}
   \frac{ r_{\bar{k}b }}   { r_{\bar{k} b'}} = \frac{\mu_{b} }{\mu_{b'}}.
\end{equation}

We refer a user who is associated with multiple BSs over the same pattern(s) as a multi-associated user. To show the number of multi-associated users is upper bounded by $B-1$, we follow the same argument as in \cite{Gajic2009} using a bipartite graph representation (BGR), where the multiple-associated users and BSs are denoted by nodes, and a edge is placed between a user and a BS to represent the association. We next show that the BGR contains a loop with zero probability.

\begin{figure*}[t]
\centering
\includegraphics[width=4.5In]{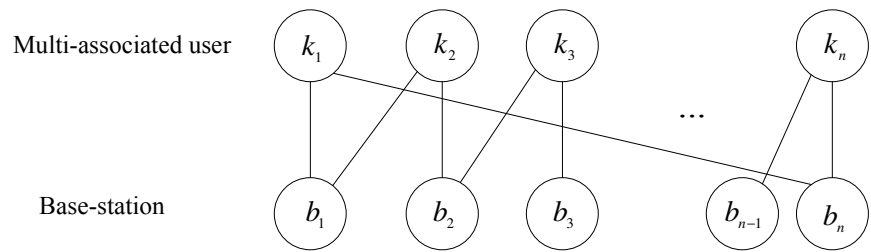}
\caption{A loop in bipartite graph representation.}\label{BGR}
\end{figure*}

Suppose there exists a loop in BGR, as shown in Fig.~\ref{BGR}. Then a sequence of nodes $k_1,b_1$,$k_2,b_2$,$\cdots$,$k_n,b_n,k_1$ exists, where nodes are different otherwise we can find a smaller loop inside. According to (\ref{cascadedEqu}), the above loop implies
\begin{equation*}
  \frac{r_{k_1 b_n }}{r_{k_1 b_1 }}  \frac{r_{k_2 b_1 }}{r_{k_2 b_2 }} \cdots \frac{ r_{k_{n} b_{n-1} }}{r_{k_n  b_n }}   =  \frac{\mu_{b_n }}{\mu_{b_1 }} \frac{\mu_{b_1 }}{\mu_{b_2 }} \cdots  \frac{\mu_{b_{n-1} }}{\mu_{b_n }} = 1,
\end{equation*}
which is obtained with zero probability. This is because $\frac{r_{k_1 b_n }}{r_{k_1 b_1 }}  \frac{r_{k_2 b_1 }}{r_{k_2 b_2 }} \cdots \frac{ r_{k_{n} b_{n-1} }}{r_{k_n  b_n }} $ is a function of independent continuous random variables, resulting in a continuous random variable itself. The probability of equaling a constant is zero.

 With this in mind, we construct the BGR by adding on one multi-associated user after the other. It is clear that the number of multi-associated users added can not be greater than the number of BSs, otherwise the bipartite graph can not be constructed without a loop, which proves the proposition.

\begin{thebibliography}{10}
\providecommand{\url}[1]{#1}

\bibitem{Fooladivanda2013}
D.~Fooladivanda and C.~Rosenberg, ``Joint resource allocation and user
  association for heterogeneous wireless cellular networks,'' \emph{{IEEE}
  Trans. Wireless Commun.}, vol.~12, no.~1, pp. 248--257, 2013.

\bibitem{Boudreau2009}
G.~Boudreau, J.~Panicker, N.~Guo, R.~Chang, N.~Wang, and S.~Vrzic,
  ``Interference coordination and cancellation for {4G} networks,''
  \emph{{IEEE} Commun. Mag.}, vol.~47, no.~4, pp. 74--81, 2009.

\bibitem{Saquib2013}
N.~Saquib, E.~Hossain, and D.~I. Kim, ``Fractional frequency reuse for
  interference management in {LTE}-advanced HetNets,'' \emph{{IEEE} Wireless
  Commun. Mag.}, vol.~20, no.~2, pp. 113--122, 2013.


\bibitem{Holfeld2014}
B.~Holfeld, A.~Relitz, T.~Wirth, and E.~Jorswieck, ``Joint multicell subchannel
  assignment with interference control and resource fairness in multiband
  {OFDMA} cellular networks,'' in \emph{IEEE International Symposium on Dynamic Spectrum Access Networks
  (DYSPAN),}, 2014, pp. 467--476.


\bibitem{Lopez-Perez2012b}
D.~Lopez-Perez, X.~Chu, and J.~Zhang, ``Dynamic downlink frequency and power
  allocation in {OFDMA} cellular networks,'' \emph{{IEEE} Trans. Commun.},
  vol.~60, no.~10, pp. 2904--2914, 2012.

\bibitem{Chang2009}
R.~Chang, Z.~Tao, J.~Zhang, and C.-C. Kuo, ``A graph approach to dynamic
  fractional frequency reuse (FFR) in multi-cell {OFDMA} networks,'' in
  \emph{IEEE ICC}, 2009,
  pp. 1--6.

\bibitem{Novlan2010}
T.~Novlan, J.~Andrews, I.~Sohn, R.~Ganti, and A.~Ghosh, ``Comparison of
  fractional frequency reuse approaches in the {OFDMA} cellular downlink,'' in
  \emph{IEEE GLOBECOM}, 2010,
  pp. 1--5.

\bibitem{Dotzler2010}
A.~Dotzler, W.~Utschick, and G.~Dietl, ``Fractional reuse partitioning for
  {MIMO} networks,'' in \emph{IEEE GLOBECOM}, 2010, pp. 1--5.

\bibitem{Ali2009}
S.~Ali and V.~Leung, ``Dynamic frequency allocation in fractional frequency
  reused {OFDMA} networks,'' \emph{{IEEE} Trans. Wireless Commun.}, vol.~8,
  no.~8, pp. 4286--4295, 2009.

\bibitem{R1-0513412005}
G.~R1-051341, ``Flexible fractional frequency reuse approach,'' in \emph{3GPP
  TSG RAN WG1 Meeting}, Seoul, Korea, 2005.

\bibitem{Yates1995}
R.~Yates and C.-Y. Huang, ``Integrated power control and base station
  assignment,'' \emph{{IEEE} Trans. Veh. Technol.}, vol.~44, no.~3, pp.
  638--644, 1995.

\bibitem{Hanly1995}
S.~Hanly, ``An algorithm for combined cell-site selection and power control to
  maximize cellular spread spectrum capacity,'' \emph{{IEEE} J. Sel. Areas
  Commun.}, vol.~13, no.~7, pp. 1332--1340, 1995.

\bibitem{Rashid-Farrokhi1997}
F.~Rashid-Farrokhi, K.~Liu, and L.~Tassiulas, ``Downlink power control and base
  station assignment,'' \emph{{IEEE} Commun. Lett.}, vol.~1, no.~4, pp.
  102--104, 1997.

\bibitem{Qian2013}
L.~P. Qian, Y.~J. Zhang, Y.~Wu, and J.~Chen, ``Joint base station association
  and power control via benders' decomposition,'' \emph{{IEEE} Trans. Wireless
  Commun.}, vol.~12, no.~4, pp. 1651--1665, 2013.

\bibitem{Ye2013}
Q.~Ye, B.~Rong, Y.~Chen, M.~Al-Shalash, C.~Caramanis, and J.~Andrews, ``User
  association for load balancing in heterogeneous cellular networks,''
  \emph{{IEEE} Trans. Wireless Commun.}, vol.~12, no.~6, pp. 2706--2716, 2013.



\bibitem{Shen2014}
K.~Shen and W.~Yu, "Distributed pricing-based user association for downlink heterogeneous cellular networks," \emph{IEEE J. Sel. Areas Commun.}, vol.~32,
  no.~6, pp. 1100--1113, 2014.

\bibitem{Madan2010}
R.~Madan, J.~Borran, A.~Sampath, N.~Bhushan, A.~Khandekar, and T.~Ji, ``Cell
  association and interference coordination in heterogeneous {LTE-A} cellular
  networks,'' \emph{IEEE J. Sel. Areas Commun.}, vol.~28,
  no.~9, pp. 1479--1489, 2010.

\bibitem{Guvenc2011}
I.~Guvenc, ``Capacity and fairness analysis of heterogeneous networks with
  range expansion and interference coordination,'' \emph{IEEE Communications
  Letters}, vol.~15, no.~10, pp. 1084--1087, 2011.

\bibitem{Guvenc2011a}
I.~Guvenc, M.-R. Jeong, I.~Demirdogen, B.~Kecicioglu, and F.~Watanabe, ``Range
  expansion and inter-cell interference coordination (ICIC) for picocell
  networks,'' in \emph{Proc. IEEE Vehicular Technology Conf. (VTC Fall)}, 2011,
  pp. 1--6.

\bibitem{Son2009}
K.~Son, S.~Chong, and G.~Veciana, ``Dynamic association for load balancing and
  interference avoidance in multi-cell networks,'' \emph{{IEEE} Trans. Wireless
  Commun.}, vol.~8, no.~7, pp. 3566--3576, 2009.

\bibitem{Kuang2012}
Q.~Kuang, J.~Speidel, and H.~Droste, ``Joint base-station association, channel
  assignment, beamforming and power control in heterogeneous networks,'' in
  \emph{Vehicular Technology Conference (VTC Spring), 2012 IEEE 75th}, 2012,
  pp. 1--5.

\bibitem{Ye2013a}
Q.~Ye, M.~Al-Shalash, C.~Caramanis, and J.~G. Andrews, ``On/off macrocells and
  load balancing in heterogeneous cellular networks,'' in \emph{Global
  Communications Conference (GLOBECOM), 2013 IEEE}, 2013, pp. 3814--3819.


\bibitem{Jin2013}
Y.~Jin and L.~Qiu, ``Joint user association and interference coordination in
  heterogeneous cellular networks,'' \emph{{IEEE} Commun. Lett.}, vol.~17,
  no.~12, pp. 2296--2299, 2013.

\bibitem{Son2011}
K.~Son, Y.~Yi, and S.~Chong, ``Utility-optimal multi-pattern reuse in
  multi-cell networks,'' \emph{{IEEE} Trans. Wireless Commun.}, vol.~10, no.~1,
  pp. 142--153, 2011.

\bibitem{Kuang}
Q.~Kuang, ``Joint user association and reuse pattern selection in heterogeneous
  networks,'' presented at \emph{IEEE the Eleventh International Symposium on Wireless Communication Systems (ISWCS 2014)}, Barcelona, Spain, August 26-29, 2014.

\bibitem{Pedersen2011}
K.~Pedersen, F.~Frederiksen, C.~Rosa, H.~Nguyen, L.~Garcia, and Y.~Wang,
  ``Carrier aggregation for {LTE}-advanced: functionality and performance
  aspects,'' \emph{{IEEE} Commun. Mag.}, vol.~49, no.~6, pp. 89--95, 2011.


\bibitem{Bazaraa2013}
M.~S. Bazaraa, H.~D. Sherali, and C.~M. Shetty, \emph{Nonlinear programming:
  theory and algorithms}, 3rd~ed.\hskip 1em plus 0.5em minus 0.4em\relax New
  York: Wiley-Interscience, 2006.

\bibitem{Raman2005}
C.~Raman, R.~Yates, and N.~B. Mandayam, ``Scheduling variable rate links via a
  spectrum server,'' in \emph{New Frontiers in Dynamic Spectrum Access
  Networks, 2005. DySPAN 2005. 2005 First IEEE International Symposium on},
  2005, pp. 110--118.

\bibitem{Bertsekas1999}
D.~P. Bertsekas, \emph{Nonlinear Programming}.\hskip 1em plus 0.5em minus
  0.4em\relax Athena Scientific, Belmont, USA, 1999.

\bibitem{Jaggi2013}
M.~Jaggi, ``Revisiting frank-wolfe projection-free sparse convex
  optimization,'' in \emph{Proceedings of the 30th International Conference on
  Machine Learning (ICML-13)}, 2013, pp. 427--435.

\bibitem{Boyd2004}
S.~Boyd and L.~Vandenberghe, \emph{Convex Optimization}.\hskip 1em plus 0.5em
  minus 0.4em\relax Cambridge University Press, New York, USA, 2004.

\bibitem{3GPP2010}
3GPP, ``Further advancements for E-UTRA physical layer aspects (TR 36.814),''
  vol. v9.0.0, 2010.

\bibitem{Deb2014}
S.~Deb, P.~Monogioudis, J.~Miernik, and J.~Seymour, ``Algorithms for enhanced
  inter-cell interference coordination (eICIC) in {LTE} HetNets,''
  \emph{{IEEE/ACM} Trans. Netw.}, vol.~22, no.~1, pp. 137--150, 2014.

\bibitem{Gajic2009}
V.~Gajic, J.~Huang, and B.~Rimoldi, ``Competition of wireless providers for
  atomic users: Equilibrium and social optimality,'' in \emph{Communication,
  Control, and Computing, 2009. Allerton 2009. 47th Annual Allerton Conference
  on}, 2009, pp. 1203--1210.

\bibitem{AndrewsJ2014}
J.G.~Andrews, S.~Singh, Q.~Ye, X.~Lin, and H.~Dhillon, "An overview of load balancing in HetNets: old myths and open problems," \emph{Wireless Communications, IEEE}, vol.21, no.2, pp.18--25, 2014

\bibitem{cvx}
CVX Research, Inc., "{CVX}: Matlab software for disciplined convex programming, version 2.0," \url{http://cvxr.com/cvx}, April 2011.

\bibitem{SinghS2013}
S.~Singh, H.S.~Dhillon, J.G.~Andrews, "Offloading in Heterogeneous Networks: Modeling, Analysis, and Design Insights," \emph{{IEEE} Trans. Wireless Commun.}, vol.12, no.5, pp.2484--2497, May 2013.

\bibitem{wsa2015}
Q.~Kuang, W.~Utschick, and A.~Dotzler, ``Multi-pattern resource allocation in heterogeneous cellular networks,'' \emph{19th International ITG Workshop on Smart Antennas (WSA 2015)}, Ilmenau, Germany, March 3-5, 2015.

\bibitem{Binnan2015}
B.~Zhuang, D.~Guo, and M.L.~Honig, "Traffic-driven spectrum allocation in heterogeneous networks," \emph{IEEE J. Sel. Areas Commun.}, vol.~33,
  no.~10, pp. 2027--2038, 2015.


\bibitem{Luo1}
M.~Hong, R.~Sun, H.~Baligh, and Z.Q.~Luo, "Joint base station clustering and beamformer design for partial coordinated transmission in heterogeneous networks, " \emph{IEEE J. Sel. Areas Commun.}, vol.~31,
  no.~2, pp. 226--240, 2013.

 \bibitem{Luo2}
 M.~Sanjabi, M.~Razaviyayn, Z.Q.~Luo, "Optimal joint base station assignment and beamforming for heterogeneous networks," \emph{IEEE Trans. Signal Process.}, vol.62, no.8, pp.1950-1961, 2014.

 \bibitem{Luo3}
 R. Sun, M. Hong, and Z.Q.Luo, "Joint downlink base station association and power control for maxmin fairness: Computation and complexity," \emph{IEEE IEEE J. Sel. Areas Commun.}, vol.33, no.6, pp.1040-1054, June 2015.

\end{thebibliography}


\end{document}